\documentclass[twocolumn,usenatbib]{mn2e}
\usepackage{aas_macros}
\usepackage{subfigure}
\usepackage{longtable}
\usepackage{array}
\usepackage{booktabs}
\usepackage{graphicx,natbib,rotating}

\title[Multi-frequency VLBA Polarimetry of the high-redshift GPS Quasar OQ172]
{Multi-frequency VLBA Polarimetry of the high-redshift GPS Quasar
OQ172}
\author[Liu et al.]
{Yi Liu $^{1,3}$\thanks{E-mail: liuyi@pmo.ac.cn}, D. R.
Jiang$^{2,3}$, Minfeng Gu$^{2}$, L. I. Gurvits$^{4,5}$
\\
$^{1}$ Purple Mountain Observatory, Chinese Academy of Sciences,
Nanjing 210008, China
\\
$^{2}$ Key Laboratory for Research in Galaxies and Cosmology,
Shanghai Astronomical Observatory, \\
Chinese Academy of Sciences, 80 Nandan Road, Shanghai 200030, China
\\
$^{3}$ Key Laboratory of Radio Astronomy, Chinese Academy of
Sciences
\\
$^{4}$ Joint Institute for VLBI ERIC, P.O. Box 2, 7990 AA Dwingeloo,
The Netherlands
\\
$^{5}$ Department of Astrodynamics and Space Missions, Delft
University of Technology, Kluyverweg 1, \\
2629 HS Delft, The Netherlands}

\begin{document}
\pagerange{\pageref{firstpage}--\pageref{lastpage}} \pubyear{2002}
\maketitle
\label{firstpage}
\begin{abstract}
Multi-frequency Very Long Baseline Array (VLBA) polarimetry
observation of the GHz-Peaked Spectrum (GPS) quasar OQ172
(J1445+0958) has been performed at 1.6, 2.2, 4.8, 8.3 and 15.3 GHz
in 2005. Core-jet structures are detected in all bands with the jet
strongly bent at about 3 mas from the core. The radio emission of
the source is polarised at all five bands. We study the Faraday
Rotation in the core and jet components at all five bands, and find
good linear fits of Faraday Rotation in the core and jet components
at 4.8 and 8.3 GHz. At these two frequencies, the Rotation Measure
(RM) is $\sim 2000~\rm rad~m^{-2}$ in the core and $\sim 700~\rm
rad~m^{-2}$ in the inner jet components and continues to decrease at
the outer jet parts. We find that the depolarisation at 4.8 and 8.3
GHz might be caused by the internal medium in the source. We
investigate consistency of the turnover spectra of VLBI components
with the Synchrotron Self-Absorption (SSA) and Free-Free Absorption
(FFA) models. Although these two models can not be easily
distinguished due to the lack of low-frequency data, the physical
parameters can be constrained for each model. We find that the large
width of the $\rm [OIII]_{5007}$ line is likely caused by a jet
interaction with a Narrow Line Region (NLR) medium. The jet bending,
significant RM variations, Faraday depolarisation, spectral
turnover, and broad line width of $\rm [OIII]_{5007}$ could be
closely related, likely caused by the same nucleus medium,
presumably NLR.


\end{abstract}


\section{Introduction}
GHz-Peaked Spectrum (GPS) radio sources are characterised by their
compact size (L $< \sim$ 100 pc) and convex spectra peaked at $\sim$
0.5-10 GHz received frequencies (O'Dea et al.\ 1991). GPS radio
galaxies often have a compact symmetric structure on the scale of
milli- to centi-arcseconds, while GPS quasars appear with a typical
core-jet morphology, in line with the Active Galactic Nucleus (AGN)
unified schemes. Like most GPS sources, the quasar OQ172
(J1445+0958) at $z=3.552$, shows no kpc-scale radio structure. VLBI
images have revealed a compact N-S elongation at 18 cm (Dallacasa et
al.\ 1995) and 6 cm (Fomalont et al.\ 2000), also detected with the
Space VLBI VSOP at 6 cm (Gurvits 2003). Two more distant compact
southern components were detected at 13 and 4 cm (Fey \& Charlot
2000). The known so far centi-arcseconds structural pattern
resembles a `clockwise semicircle' with the core in the northernmost
position.

At the peak `turnover' frequency of GPS sources, $\nu_{T}$,
optically thin radio synchrotron radiation (for $\nu>\nu_{T}$)
transits to optically thick emission (for $\nu<\nu_{T}$), though
there is still a debate over whether this is caused by Synchrotron
Self-Absorption (SSA) or Free-Free Absorption (FFA) (O'Dea 1998).
O'Dea \& Baum (1997) and Snellen et al.\ (2000) suggested that the
observed anti-correlation between $\nu_{T}$ and the linear size of
GPS sources could be explained by SSA. However, Bicknell et al.\
(1997) proposed an alternative model with FFA occurring in an
ionized gas surrounding the radio emitting regions. To discriminate
between these two models, multi-frequency VLBI studies are required.


OQ172 has an extremely high rest-frame rotation measure (RM $>$
20,000 $\rm rad~m^{-2}$; Udomprasert et al.\ 1997), one of the
highest among $\sim$ 20 known high-RM sources. At only 10 mas
(R$_{proj}\sim$ 90 pc) from the VLBI core, the RM falls to $<$ 100
$\rm rad~m^{-2}$. 
Such a steep RM gradient might cause intrinsic depolarisation of the
synchrotron emission in the VLBI core regions.
The former could be caused by weakening of the regular magnetic
field and/or increasing density of the ionised gas.

As a further evidence of unusual properties of the physical
environment in the core of OQ172, the near infrared (NIR) spectrum
shows a typical H$\beta$ broad emission line of Full Width at Half
Maximum (FWHM) of $\sim$ 3,700 $\rm km~s^{-1}$, but an unusually
large width of 2,200 $\rm km~s^{-1}$ for the narrow $\rm
[OIII]_{5007}$ line (Hirst et al.\ 2003). Mantovani et al. (1998)
relate the width of the $\rm [OIII]_{5007}$ line to the interaction
of the strongly bent VLBI jet with the NLR medium.

We performed a 10-hour-long multi-frequency VLBA polarimetry
observation at 1.6, 2.2, 4.8, 8.3 and 15.3 GHz aiming to investigate
the unusual physical circumnuclear environment of the GPS quasar
OQ172 by studying: (1) the distribution of polarised emission and
rotation measure, (2) the GPS absorption mechanism and associated
opacity, and (3) variations of physical parameters along the jet.
Specifically, we aimed to estimate parameters of the circumnuclear
environment of OQ172 on the basis of the observed properties at the
redshifted rest-frame frequencies within the range 6-70 GHz.

The Hubble constant $H_{0}=70 \rm {~km ~s^ {-1}~Mpc^{-1}}$,
$\Omega_{M}=0.3$ and~ $\Omega_{\Lambda} = 0.7$ are used throughout
this paper. With these cosmological parameters, at the redshift of
OQ172, 1 mas corresponds to 8.84 pc. The paper is organised as
follows. In Sect. 2 the observation and data reduction are
described. The results and discussions are presented in Sect. 3,
followed by our conclusions in Sect. 4.

\section{Observation and Data Reduction}
Our multi-frequency polarimetric continuum observation of OQ172 was
conducted at 1.6, 2.2, 4.8, 8.3, and 15.3 GHz on 2005 June 22 with
the VLBA for the total duration of about 10 hours. The left- and
right-circular polarisation (LCP and RCP, respectively) signals were
recorded in 8 baseband channels (BBCs) with a total bandwidth of 128
MHz and 2-bit sampling. BBCs were placed in close pairs to avoid
Faraday depolarisation across each pair for an assumed rest-frame RM
$<$ 50,000 $\rm rad~m^{-2}$. In each frequency band, the BBCs were
separated as widely as possible in order to achieve the most
accurate estimate of the rotation measure. The theoretical RMS noise
in total intensity images with integration time of 20 mins at 1.6,
2.2, 4.8 GHz, 90 mins at 8.3 GHz and 160 mins at 15.3 GHz do not
exceed 0.15 $\rm mJy~beam^{-1}$. However, the real RMS noise
measured at each band is usually higher than the theoretical value,
which is shown in Figure 2. In each of the 5 frequency bands, we
observed the primary flux density calibrator 3C286 in 2 scans (4 min
each), the instrumental polarisation ($D$-term) calibrator OQ208
over PA range $>100^{\circ}$ in 5 scans (6 min each). Three scans
were used for each of two electric vector position angle (EVPA)
calibrators, 3C279 (3 min each) and 3C345 (4 min each). Table 1
lists the details of the observation.

\begin{table*}
\caption{VLBA observational parameters} \label{table2}
\begin{tabular}{lcccccccccc}
\hline\hline Sources& IF Frequencies (GHz)& BandWidth(MHz) & Scan &
Total on-source integration (min)\\
               (1) &  (2)     & (3)& (4)  &  (5)\\
\hline
OQ172, OQ208, 3C286 & 1.412,     1.502,      1.592,      1.683 &        8 &      24 &          72 \\
OQ172, OQ208, 3C286 & 2.166,     2.211,      2.256,      2.302 &        8 &      24 &          72 \\
OQ172, OQ208, 3C286 & 4.630,     4.770,      4.910,      5.051 &        8 &      24 &         102 \\
OQ172, OQ208, 3C279 & 8.116,     8.196,      8.396,      8.537 &        8 &      23 &         134 \\
OQ172, OQ208, 3C279 &15.140,    15.220,     15.421,     15.561 &        8 &      34 &         191 \\
\hline\hline
\end{tabular}
\end{table*}

The recorded data were first processed at the VLBA correlator in
Socorro (New Mexico, USA) and then calibrated and fringe-fitted
using the NRAO Astronomical Image Processing System (AIPS) software,
as described by Zavala \& Taylor (2004). Self-calibration and
imaging were with the Difmap program (Shepherd, Pearson, \& Taylor
1994). The estimated uncertainty of the amplitude calibration in
self-calibration process is about $5\sim10$ percent, these values
are consistent with typical amplitude corrections reported by Homan
et al. 2002. In the course of the data reduction, each of four
different IF channels at each of the five observing bands was
processed separately. In all steps of the data processing, we used
the VLBA Los Alamos as the reference antenna. We used the
non-polarised compact source OQ208 to calibrate the instrumental
polarisation of the antennas by using the AIPS task LPCAL. To ensure
that the $D$-term solutions were physically meaningful, we checked
the distribution of the normalized RL/RR visibility in complex (real
and imaginary) plane, with and without instrumental calibration. We
found that the RL/RR visibility appeared well clustered after the
instrumental calibration (see
http://www.vla.nrao.edu/astro/calib/polar/). The solution shows that
the instrumental polarisations are typically about 1\% at 1.6, 2.2,
4.8 GHz and about 1.5\% for 8.3 and 15.3 GHz.

Since EVPA calibrators recommended by NRAO are all stable across
bandwidth (see http://www.vla.nrao.edu/astro/calib/polar/), we
calibrated EVPA at each baseband by comparing to the calibrator at a
close arbitrary frequency assuming small rotation across base-bands.
Since the source 3C286 is a stable EVPA calibrator with its R/L
phase difference of $66^{\circ}$ (see
http://www.vla.nrao.edu/astro/calib/polar/), we used this source to
correct the EVPA of the target source at 1.6, 2.2, and 4.8 GHz. The
absolute EVPA calibration at 8.3 GHz was performed by using an
integral EVPA of 3C279 within 6 days (observed on 2005 Jun 17) of
our observation from the NRAO VLA/VLBA Polarisation Calibration
Database, in which the R/L phase difference was given as
$-142.2\pm1.8^\circ$. However, 3C279 has no available absolute EVPA
at 15.3 GHz, thus the true EVPA at this frequency was estimated by
using RM of $-79~\rm rad~m^{-2}$ obtained at the two lower
frequencies of 4.8 and 8.3 GHz (Chen, Shen \& Feng 2010). This led
us to the estimate of the R/L phase difference for 3C279 at 15.3 GHz
of $112.5\pm 1.7^\circ$.

\section{RESULTS AND DISCUSSION}
\subsection{The Radio Structure}
Since the steep-spectrum structures fade away rapidly with increased
frequencies, the size of detected radio emission in an intensity
distribution image decreases rapidly with increased frequencies.
This physical effect is further amplified by typically lower
sensitivities of VLBI systems at higher frequencies. The total
intensity images of OQ172 observed at five frequencies have been
presented as preliminary results in our previous paper (Liu et al.
2016). In this paper, we update the images (Figure 1) by adding the
model-fit on individual components (core and jet, see below). The
figure clearly demonstrates a consistent core-jet morphology at all
five bands (see also Liu et al. 2016). A combination of all five
images offers an opportunity to trace the jet in details. Our 1.6
and 4.8 GHz images are consistent with those by Udomprasert et al
(1997). At 2.2 GHz, the source has a very similar structure to that
at 1.6 GHz, albeit more resolved in outer jet regions. At 8.3 and
15.3 GHz, the images show jet bending at $\sim$ 3 mas from the core.
At the higher frequencies of 8.3 and 15.3 GHz and correspondingly
higher angular resolution, the jet knots are fragmented into smaller
size features. A combination of our images at five frequency shows
the overall jet structure as pointed toward the west from the core,
followed by a sharp bend toward south at about 3 mas and yet another
bend toward east at about 20 mas.

After obtaining the source's images, we performed model-fitting on
the self-calibrated visibility data with circular Gaussian
components. We assumed the brightest component in each image to be
stationary and used it as a reference point. The coordinates of
other components were measured relative to it. Figure 1 shows the
naturally weighted images with model-fitted circular Gaussian
components at each band. Parameters of the fitted components and the
derived brightness temperature ($T_{B}$) in the source frame for
both components are listed in  Table 2. The source-frame $T_{B}$ is
estimated by using the following expression (Ghisellini et al.
1993),
\begin{equation}\label{eq1}
T_B=1.77\times10^{12}(\frac{f_\nu}{\rm{Jy}})(\frac{\nu}{\rm{GHz}})^{-2}(\frac{\theta_d}{\rm{mas}})^{-2}(1+z),
\end{equation}
where $f_\nu$ is flux density at frequency $\nu$, and $\theta_d$ is
the component angular size.

\subsection{The Polarisation Properties and Rotation Measure}
In Figure 2, we show the integrated polarisation images, which
include all baseband channels at each frequency. We found that the
polarisation images at a single baseband are similar to the
integrated ones at all five bands. This implies that the bandwidth
depolarisation is generally not significant. The polarisation
structures of OQ172 are detected in all five frequency bands. The
region with a detectable linear polarisation decreases gradually in
size from 20 mas to a few mas with increasing observational
frequencies due to increasing resolution and lower surface
brightness. Compared to `normal' AGNs with typical radio
polarisation at the level of several percents, GPS sources are
generally less polarised (O'Dea 1998; Stanghellini et al. 1998). The
overall fractional polarisation from low to high frequencies in our
observation is 0.39\%, 0.46\%, 1.60\%, 1.38\% and 0.6\% in five
frequency bands, respectively. The fractional polarisation is higher
in the jet components than that in the core, as clearly visible in
Figure 2. The relatively low polarisation level in the core region
is typical for radio-loud AGNs and might be due to the inner regions
effects of high opacity, high depolarisation, complex morphology and
plasma effects in the immediate vicinity of AGN (e.g. Taylor 1998;
Homan et al. 2009, Chen et al. 2010). The depolarisation effect
across the 8 MHz sub-band is generally small at all five
frequencies, however, the relatively low fractional polarisation at
two low frequencies (1.6 and 2.2 GHz) could still be partly caused
by the depolarisation from the Faraday rotation, which is expected
to be higher within the same width of a sub-band compared to the
higher frequencies. A lower resolution at a lower frequency adds up
to this effect (Chen et al. 2010; 2011).

The polarisation structure at 4.8, 8.3 and 15.3 GHz can be traced
with higher degree of detail than at lower frequencies of 1.6 and
2.2 GHz due to an in-beam depolarisation at a lower angular
resolution. Comparing to the earlier 5 GHz polarisation image
presented by Udomprasert et al. (1997), we found that the
polarisation level in the core region is lower although the two
observations provided the similar total flux density. In order to
search for possible variability in polarised structures in OQ172, we
have re-processed the Udomprasert's data from the NRAO archive,
using the same processing approach as for our own data. In order to
evaluate a potential effect of bandwidth depolarisation, we compared
the polarised images at each sub-band by Udomprasert's (1997) with
those from our data. We find that the fractional polarisation in the
core component at each sub-band is all lower than that in
Udomprasert et al. (1997). This, together with the similar setting
of segregated sub-bands in Udomprasert's data and ours, indicates
strongly that the lower fractional polarised flux density in the
core component in our 4.8 GHz data is unlikely due to the bandwidth
depolarisation effect. We also reduced another VLBA archive data set
from the epoch 2000.6 at 5 GHz. Figure 3 shows the distributions of
visibility amplitudes of OQ172 at these three epoches. We found an
obvious change of visibility amplitudes at long baselines. The
visibility amplitude was almost 20 times at the second epoch 2000.6
at the $uv$ radius longer than $90 \times 10^{6} \lambda$. The
visibility amplitudes remained high at the longest baselines in the
third available observing epoch 2005.6 too. While the new component
was not found close to the core in our and previous 5 GHz
observations, a new component can be clearly identified in our 15.3
GHz image (component U7, see Figure 1). This suggests that there
would has been a new component when the visibility amplitude changes
but it needs higher resolution to distinguish. The new component
might manifest itself in both the source's flux density and
polarisation properties. This new component does not increase the
total flux density, therefore, it could be a newly-seen but not
new-born component in the innermost core region, moving outwards
relative to the core. It might affect the polarisation flux and lead
to a variation of the distribution of
the polarised emission at 4.8 GHz. 

The linear polarisation can be estimated for each component by using
the total intensity and the Stokes $Q$ and $U$ flux densities from
the modelfitting. The corresponding linearly polarised flux density
and positional angle were obtained as $P=(Q^{2}+U^{2})^{1/2}$ and
$\chi=1/2 \arctan(U/Q)$, respectively. The fractional polarisation
$m$ is calculated as $m=P/I$.  The uncertainties of $P$, $\chi$ and
$m$ were derived using a method suggested by Hovatta et al. (2012),
which includes the $D$-term leakage effect. The estimates of $P$,
$\chi$, and $m$ are listed in Table 3 for each component in four IFs
at all five frequencies, as well as their uncertainties.

The rotation of the polarisation plane in the magnetised medium is
given by the following formula (e.g. Spitzer 1978; Taylor 1998):
\begin{equation}
RM=812 \int_{0}^{L} n_{e}B_{\parallel}dl~~\rm rad~m^{-2},
\end{equation}
where RM is the Rotation Measure in $\rm rad~m^{-2}$, $n_{e}$ is the
electron number density in $\rm cm^{-3}$, $B_{\parallel}$ is
magnetic field in mG, $l$ is the length in pc alone the line of
sight from the observer, and $L$ is the path length in pc. As
expected, we found that the polarisation angles $\chi$ were
different at different frequencies. There are significant variations
in polarisation angles even in adjacent IF channels within the same
frequency band, while the variations in polarised flux densities are
less pronounced.

\begin{table*} \caption{{\large Component parameters in
OQ172.}}
\begin{tabular}{lccccccccccc}
\hline\hline Frequency & Component  & Flux Density & Radius from Core & Position Angle  & Size & Brightness Temperature \\
(GHz) &  & (Jy) & (mas) & (deg) & (mas) &($10^{10}$~K)  \\
              (1) & (2) & (3) & (4) & (5) & (6) & (7)  \\
\hline
1.6 GHz   & Core   &0.766$\pm$0.074  &     0.00                &    ... &  1.93  &  69.78   \\
          & L5     &0.492$\pm$0.064  &     3.38$\pm$  0.04     &   -148 &  1.60  &  63.56   \\
          & L4     &0.381$\pm$0.053  &    11.06$\pm$  0.09     &   -174 &  2.37  &  22.43   \\
          & L3     &0.174$\pm$0.041  &    14.58$\pm$  0.10     &   -178 &  2.29  &  10.97   \\
          & L2     &0.054$\pm$0.020  &    21.71$\pm$  0.74     &   -199 &  4.85  &   0.76   \\
          & L1     &0.054$\pm$0.020  &    24.34$\pm$  0.43     &   -215 &  3.10  &   1.86   \\
\hline
2.2 GHz   & Core   &0.859$\pm$0.081  &      0.00               &    ... &  2.02  &  33.12   \\
          & S5     &0.382$\pm$0.058  &     3.12$\pm$  0.05     &   -140 &  1.40  &  31.39   \\
          & S4     &0.236$\pm$0.044  &     9.12$\pm$  0.14     &   -172 &  2.50  &   6.08   \\
          & S3     &0.207$\pm$0.041  &    14.04$\pm$  0.11     &   -175 &  1.75  &  10.89   \\
          & S2     &0.042$\pm$0.025  &    20.68$\pm$  1.65     &   -196 &  6.05  &   0.18   \\
          & S1     &0.031$\pm$0.017  &    24.65$\pm$  0.60     &   -215 &  2.70  &   0.68   \\
\hline
4.8 GHz   & Core   &0.256$\pm$0.029  &      0.00               &    ... &  0.30  &  96.75   \\
          & C6     &0.234$\pm$0.028  &     1.38$\pm$  0.03     &   -111 &  0.88  &  10.28   \\
          & C5     &0.117$\pm$0.020  &     3.08$\pm$  0.03     &   -106 &  0.60  &  11.05   \\
          & C4     &0.077$\pm$0.017  &     4.32$\pm$  0.05     &   -123 &  0.87  &   3.46   \\
          & C3     &0.054$\pm$0.014  &     6.53$\pm$  0.08     &   -143 &  1.02  &   1.77   \\
          & C2     &0.067$\pm$0.019  &    10.62$\pm$  0.45     &   -165 &  3.29  &   0.21   \\
          & C1     &0.054$\pm$0.014  &    14.91$\pm$  0.22     &   -170 &  1.96  &   0.48   \\
\hline
8.3 GHz   & Core   &0.304$\pm$0.028  &      0.00               &    ... &  0.31  &  36.67   \\
          & X6     &0.124$\pm$0.019  &     1.32$\pm$  0.05     &   -111 &  0.93  &   1.66   \\
          & X5     &0.074$\pm$0.014  &     3.18$\pm$  0.07     &   -109 &  0.89  &   1.08   \\
          & X4     &0.032$\pm$0.010  &     4.76$\pm$  0.16     &   -127 &  1.24  &   0.24   \\
          & X3     &0.019$\pm$0.008  &     6.86$\pm$  0.16     &   -143 &  0.98  &   0.23   \\
          & X2     &0.039$\pm$0.025  &    11.05$\pm$  1.27     &   -164 &  3.92  &   0.03   \\
          & X1     &0.016$\pm$0.008  &    14.72$\pm$  0.34     &   -169 &  1.49  &   0.08   \\
\hline
15.3 GHz  & Core   &0.181$\pm$0.025  &      0.00               &    ... &  0.30  &   6.85   \\
          & U7     &0.036$\pm$0.013  &      .51$\pm$  0.02     &   -142 &  0.29  &   1.46   \\
          & U6     &0.035$\pm$0.012  &     1.47$\pm$  0.08     &   -114 &  0.58  &   0.35   \\
          & U5     &0.024$\pm$0.011  &     3.18$\pm$  0.16     &   -107 &  0.77  &   0.14   \\
          & U4     &0.013$\pm$0.010  &     4.27$\pm$  0.42     &   -123 &  1.10  &   0.04   \\
          & U3     &0.008$\pm$0.010  &     6.70$\pm$  0.76     &   -140 &  1.20  &   0.02   \\
\hline \hline
\end{tabular}
\begin{quote}
\ Notes: (1): Observing frequency. (2): Component name. (3): Total
flux density of each components. (4): Distance of each component
from the core. (5): Position angle of each component. (6): Major
axis of each component. (7): the bright temperature of each
component in units of $10^{10}$ K.
\end{quote}
\end{table*}

In order to examine whether the variation of polarisation angles
($\chi$) is associated with Faraday Rotation, we attempted to obtain
linear fits over the entire wavelength range at each frequency band
allowing for $n\pi$ ambiguities in the polarisation position angles
with their uncertainties. Figures 4 - 6 present the `$\chi -
\lambda^{2}$' relation for core (in Figure 4) and jet components
(1.6 and 2.2 GHz in Figure 5 and 4.8, 8.3, and 15.3 GHz in Figure 6)
at all five frequencies. The RM measurements of the core and two
innermost jet components at 4.8 and 8.3 GHz were presented as a
preliminary results in our previous paper (Liu et al. 2013). In this
work, we reproduce those figures in order to make systematic study
by combining with other bands, and by extending it to all jet
components. At two lower frequencies (1.6 and 2.2 GHz), both the
direct measurements (in each middle panel) of polarisation position
angles and those with $n\pi$ rotational ambiguities are displayed
for the core component in Figure 4(a) and 4(b). While there is no
good linear fit on the directly measured polarisation position
angles, the rotation measure $\sim 2000~\rm rad~m^{-2}$ could be
obtained at both frequencies from the linear fit by adding minimal
$n\pi$ rotational ambiguities. However, we find that a similar RM
can always be obtained by adding same rotational ambiguities for any
given polarisation angles at four IFs. Therefore, the RM of core
component estiamted from adding rotational ambiguities at 1.6 and
2.2 GHz is not reliable. In contrast, a good linear fit without
adding any rotational ambiguities can be found for the core
component at 4.8 and 8.3 GHz, which are plotted in Figure 4(c) and
4(d). At 4.8 GHz, the RM value of the core component,
$2089\pm225~\rm rad~m^{-2}$, is very similar to that at 8.3 GHz,
$2028\pm172 ~\rm rad~m^{-2}$. At 15.3 GHz, the `$\chi -
\lambda^{2}$' relation is apparently non-linear, which likely
implies a contribution from rotation within the core itself. By
adding $n\pi$ rotational ambiguities, a RM value of
$2.911\times10^{5}\pm1082~\rm rad~m^{-2}$ is found, more than two
orders of magnitude higher than those at 4.8 and 8.3 GHz. Such an
extremely high value, and the inconsistence with other bands,
indicate that the value should be treated with caution.

In Figures 5 and 6, the `$\chi - \lambda^{2}$' relations for each
jet component at all five frequencies are presented, in which only
the directly measured polarisation position angels in each single
baseband are plotted. In most cases, there are no significant linear
relation in jet components at 1.6 (Figure 5(a)-(e)) and 2.2 GHz
(Figure 5(f)-(j)). However, the variation of $\chi$ tends to
decrease along the jet, likely indicating a smaller RM towards outer
jet regions. At 4.8 GHz (Figure 6(a)-(f)) and 8.3 GHz (Figure
6(g)-(l)), except for C3 and X3, all jet components show a good
linear fits on the `$\chi - \lambda^{2}$' relation, and the
resulting RM values of the corresponding jet components are
generally consistent within the errors at two bands. The innermost
jet component C6 and X6 show a comparable RM, $1972\pm207 ~\rm
rad~m^{-2}$ and $2102\pm124 ~\rm rad~m^{-2}$, respectively. However,
the RM values drop to $722\pm132 ~\rm rad~m^{-2}$ in C5 and
$780\pm105 ~\rm rad~m^{-2}$ in X5 at a distance of $\sim$ 3 mas from
the core.  Along the jet, the RM values show a similar tendency of
monotonic decrease at 4.8 and 8.3 GHz, from $455\pm96 ~\rm
rad~m^{-2}$ in C4 to $113\pm36 ~\rm rad~m^{-2}$ in C1, and
$470\pm127 ~\rm rad~m^{-2}$ in X4 to $176\pm138 ~\rm rad~m^{-2}$ in
X1. Our results at 4.8 GHz are consistent with that in Udomprasert
et al (1997). The RM can be obtained even in the outer jet
components at 4.8 and 8.3 GHz, i.e., C1 and X1, suggesting that the
Faraday screen extends to much larger radii in this source. Besides
consistency check, we also tried to connect the RM values at 4.8 and
8.3 GHz, and found that the core and inner three jet components
(C6/X6, C5/X5 and C4/X4) are consistent with appropriately adjusted
rotational ambiguities, but the three outer jet components are not.
This indicates that the medium responsible for Faraday rotation in
the core and inner jet regions is different from that in the outer
jet. At 15.3 GHz (Figure 6(m)-(q)), it appears that all the `$\chi -
\lambda^{2}$' relations based on the directly measured polarisation
angles are non-linear, thus reliable estimates of RM cannot be
obtained. At low frequencies (1.6 and 2.2 GHz), the rotational
ambiguities might be the main issue when obtaining a RM value from a
linear fit. However, at 15.3 GHz, the rotational ambiguities are
less of a problem. The observed non-linear `$\chi - \lambda^{2}$'
relation may be caused by the low signal to noise ratio of the
polarisation flux detection since the corresponding uncertainties
for polarisation position angles are relatively larger.
Quantitatively, a RM value of 2500 $\rm rad~m^{-2}$ corresponds to a
polarisation angle variation of about 0.053 rad (i.e., $\rm
\sim3^{\circ}$) in observing band of $\rm 0.21\times10^{-4}~m^{2}$
at 15.3 GHz. It therefore requires a very high precision in
polarisation angle measurements. The situation is even worse for a
lower RM value. A broader bandwidth is needed to the observed RM
properties can measure the RM reliably in these cases.
Alternatively, the observed RM properties can be related to the
properties of the jet itself, such as an internal rotation within
the jet.

Table 3 shows the depolarisation in the core and jet components
between different frequencies. In our observations, the bandwidth
depolarisation across the 8 MHz sub-band cannot explain the
depolarisation. While interpretation of depolarisation in the core
by opacity effects is difficult, the depolaristion of optically thin
outer jet components might indicate on internal or external origin
of the effect. A number of depolarisation models are reported in the
literatures: the Slab model (Burn 1966), Tribble model (Tribble
1991), Rossetti-Mantovani model (Rossetti et al. 2008; Mantovani et
al. 2009), and the Repolariser model (Homan et al. 2002; Mantovani
et al. 2009; Hovatta et al. 2012). In this work, we investigate
whether the depolarisation is caused by an internal or external
Faraday screen by studing the depolarisation behavior in the jet
components and by using the model described in Hovatta et al.
(2012), in which a linearized formula $\ln m = \ln
m_{0}+b\lambda^{4}$ is used for fitting the observed fractional
polarisation $m$ at the corresponding wavelength $\lambda$. This
method provides the total depolarisation $b$ from the slope of the
fit. Then, the position on the diagram of the fitted $b$ with the
observed $\rm |RM|$, will give us clues on whether the studied
depolarisation is caused by internal or external Faraday screen (see
details in Figure 7 of Hovatta et al. 2012). Here, we only focused
on 4.8 and 8.3 bands, at which the best RM linear fits were found
for most jet components. The linear fits parameters, $\rm ln~m_{0}$
and slope $b$ together with rotation measure values $\rm RM$, are
listed in Table 4. We find that the jet components with measured RM
at 4.8 and 8.3 GHz are all above the solid line in Figure 7 of
Hovatta et al. (2012), indicating that the depolarisation behavior
in the jet is dominated by an internal Faraday depolarisation at
these two frequencies, even in the outer regions of the jet.

The RM associated with the Milky Way is usually not greater than 200
$\rm rad~m^{-2}$ at any direction (Pushkarev 2001; Pshirkov et al.
2011; Oppermann et al. 2012). At the direction of OQ172, the
approximate value of Galactic RM is less than several tens of $\rm
rad~m^{-2}$. Not only is this much smaller than the RM values in the
core and inner jet components, but also much smaller than the RM
values in the outer jet components. Therefore, the major
contribution to the high RM values is associated with the medium
outside the Milky Way. Although we cannot rule out that this
contribution is provided by the intervening intergalactic medium, we
consider as more plausible an assumption that the high RM values are
formed in the immediate vicinity of OQ172. OQ172 is one of only 20
sources known for having the highest RM values in excess of 1000
$\rm rad~m^{-2}$. The rest-frame value of the RM is higher than the
observed one by the factor $(1+z)^{2}$, leading to $\sim 40000~\rm
rad~m^{-2}$ at the redshift of OQ172. About half of these highest RM
sources are Compact Steep Spectrum (CSS) sources (Taylor, Inoue, \&
Tabara 1992). The GPS sources are even smaller in size than CSS
sources, which is commonly explained by their youth and/or
confinement by the surrounding medium. Indeed, the extremely high
value of RM in the core and inner jet regions of OQ172 is indicative
of the presence of dense gas within central $\sim$ 3 mas,
corresponding to the projected linear size of about 26.5 pc.

\begin{table*}
\begin{scriptsize} \centering \caption{{\large Parameters of polarised components in
OQ172}}
\begin{tabular}{lccccccccccc}
\hline Freq (GHz) &  & Core & L5 & L4  & L3 & L2 & L1 \\
              (1) & (2) & (3) & (4) & (5) & (6) & (7) & (8) & (9)\\
\hline
1.436  &  $P~\rm (mJy)$     &   4.57 $\pm$   1.63    &   2.84  $\pm$    1.46 &   4.92 $\pm$  1.41  &   3.15 $\pm$     1.26 &   4.57 $\pm$  1.25  &   6.72 $\pm$   1.25  &   ...          \\   %
       &  $\chi~(^\circ)$   &  -88.2 $\pm$   10.2    &   -9.8  $\pm$   14.8  &   68.1 $\pm$  8.2   &  32.7  $\pm$    11.4  &  -21.3 $\pm$  7.8   &   13.4 $\pm$   5.3   &   ...           \\   %
       &  $m~(\%)$          &    0.6 $\pm$   0.2     &    0.5  $\pm$    0.3  &   1.3  $\pm$  0.3   &   2.8  $\pm$     1.3  &    8.5 $\pm$  3.0   &   11.3 $\pm$   1.9   &   ...           \\   
1.466  &  $P~\rm (mJy)$     &    6.17$\pm$  2.12    &    1.63 $\pm$    1.97  &   6.31 $\pm$   1.93  &   3.79 $\pm$    1.83  &    5.81$\pm$   1.83   &    7.09$\pm$    1.83&   ...           \\   %
       &  $\chi~(^\circ)$   &  -23.4 $\pm$  9.8     &   -15.5 $\pm$   17.2   &   77.4 $\pm$   8.8   &   6.7  $\pm$   13.8   &  -24.2 $\pm$   9.0    &   23.2 $\pm$    7.4 &   ...           \\   %
       &  $m~(\%)$          &    0.8 $\pm$  0.3     &    0.3  $\pm$    0.4   &   1.6  $\pm$   0.4   &   2.2  $\pm$    1.7   &   10.8 $\pm$   3.7    &   13.2 $\pm$    3.2 &   ...           \\   
1.636  &  $P~\rm (mJy)$     &    4.81$\pm$  2.05    &    1.68 $\pm$    1.81  &   3.87 $\pm$   1.77  &   4.11 $\pm$     1.70  &    6.87$\pm$   1.69   &    6.49$\pm$    1.69     &   ...          \\   %
       &  $\chi~(^\circ)$   &  -51.1 $\pm$ 12.2     &   -50.3 $\pm$   19.5   &  56.8  $\pm$  14.2   &  26.8  $\pm$    11.9   &  -20.1 $\pm$   7.0    &  40.8 $\pm$    7.4     &   ...          \\   %
       &  $m(\%)$           &    0.6 $\pm$  0.3     &   0.4   $\pm$    0.4   &   1.0  $\pm$   0.5   &   2.4  $\pm$     0.9   &   12.1 $\pm$   2.9    &   13.5 $\pm$    3.5     &   ...          \\   
1.687  &  $P~\rm (mJy)$     &    3.48 $\pm$  2.16   &   1.93  $\pm$    1.93  &   3.67 $\pm$   1.90  &   3.77 $\pm$     1.83  &    5.74$\pm$   1.82   &    6.34$\pm$    1.82     &   ...          \\   %
       &  $\chi~(^\circ)$   &   -73.5 $\pm$  17.8   &  -7.4   $\pm$   17.1   &   68.9 $\pm$  15.5   &  44.2  $\pm$    10.9   &  -14.0 $\pm$   9.0    &   29.0 $\pm$    8.2     &   ...          \\   %
       &  $m~(\%)$          &    0.4  $\pm$  0.3    &   0.4   $\pm$    0.4   &   0.4  $\pm$   0.4   &   2.7  $\pm$     1.1   &   10.6 $\pm$   2.6    &   12.8 $\pm$    5.0     &   ...          \\   
\hline Freq (GHz) &  & Core & S5  & S4 & S3 & S2 & S1\\                                                                                                                             %
              (1) & (2) & (3) & (4) & (5) & (6) & (7) & (8) & (9) \\                                                                                                                         %
2.170  &  $P~\rm (mJy)$     &    4.72$\pm$     2.95   &   1.77 $\pm$    2.97   &   8.17 $\pm$  2.92  &  13.84 $\pm$    2.91 &    6.13  $\pm$  2.88  &    3.41 $\pm$   2.88  &   ...           \\  %
       &  $\chi~(^\circ)$   &   162.7 $\pm$   17.9    & -20.7  $\pm$   19.4    & 111.2  $\pm$ 10.2   &  15.1  $\pm$    6.0  &  15.4    $\pm$ 13.4   &   90.5  $\pm$  24.1   &   ...           \\  %
       &  $m~(\%)$          &    0.5 $\pm$     0.6    &   0.5  $\pm$    0.5    &   2.2  $\pm$  0.8   &  6.7   $\pm$    0.8  &   13.6   $\pm$  6.3   &   13.1  $\pm$  11.0  &   ...           \\  
2.200  &  $P~\rm (mJy)$     &    5.57$\pm$     2.40   &   1.17 $\pm$    2.19   &   7.87 $\pm$  2.14  & 12.34  $\pm$    2.14 &    6.25  $\pm$  2.12  &    3.11 $\pm$   2.12  &   ...           \\  %
       &  $\chi~(^\circ)$   &   104.8 $\pm$   12.3    & -30.5  $\pm$   19.4    &   81.2 $\pm$ 7.7   & 5.7    $\pm$    4.9  &  10.5    $\pm$  9.7   &   85.5  $\pm$  19.4  &   ...           \\  %
       &  $m~(\%)$          &    0.6 $\pm$     0.3    &   0.3  $\pm$    0.5    &   2.1  $\pm$  0.9   &  5.9   $\pm$    0.9  &   14.5   $\pm$  5.2   &   11.7  $\pm$   6.6  &   ...           \\  
2.245  &  $P~\rm (mJy)$     &   5.45 $\pm$     2.06   &   2.65 $\pm$    1.76   &   7.92 $\pm$  1.71  & 13.29  $\pm$    1.70 &     6.69 $\pm$  1.68  &    3.89 $\pm$   1.68  &   ...           \\  %
       &  $\chi~(^\circ)$   &   170.2$\pm$    10.8    & -57.0  $\pm$   18.9    &  95.9  $\pm$  6.1   &-35.4   $\pm$    3.6  &    11.5  $\pm$  7.1   &  88.9   $\pm$  12.3  &   ...           \\  %
       &  $m(\%)$           &   0.6  $\pm$     0.2    &   0.7  $\pm$    0.4    &   2.7  $\pm$  0.7   &  6.4   $\pm$    0.8  &    15.9  $\pm$  3.9   &   12.5  $\pm$   5.4  &   ...           \\  
2.306  &  $P~\rm (mJy)$     &   3.98 $\pm$     2.71   &   1.25 $\pm$    2.49   &   8.38 $\pm$  2.45  & 12.34  $\pm$    2.45 &     5.18 $\pm$  2.43  &    4.78 $\pm$   2.43  &   ...           \\  %
       &  $\chi~(^\circ)$   &  93.3  $\pm$    19.4    &  -52.1 $\pm$   17.2    &  74.0  $\pm$  8.3   &-24.7   $\pm$    5.6  &    10.1  $\pm$ 13.4   &  85.2   $\pm$  14.5  &   ...          \\  %
       &  $m~(\%)$          &   0.5  $\pm$     0.3    &   0.3  $\pm$    0.6    &   2.8  $\pm$  1.0   &  6.0   $\pm$    1.1  &     12.3 $\pm$  5.5   &   15.9  $\pm$   8.1  &   ...              \\  
\hline Freq (GHz) &  & Core & C6 & C5  & C4 & C3 & C2 & C1\\
              (1) & (2) & (3) & (4) & (5) & (6) & (7) & (8) & (9) \\
4.634  &  $P~\rm (mJy)$     &   0.90 $\pm$     0.55   &   1.06 $\pm$      0.54  &   4.89 $\pm$  0.46   &  1.46 $\pm$    0.43 &     1.64 $\pm$    0.43 &    3.82 $\pm$   0.43  &     5.72 $\pm$   0.43   \\  %
       &  $\chi~(^\circ)$   &  97.1  $\pm$    17.3    & 122.1  $\pm$     14.5   &  89.4  $\pm$  2.6    & 51.5  $\pm$    8.5  &    15.0  $\pm$    7.4  &   77.3  $\pm$   3.2   &    28.8  $\pm$   2.1   \\  %
       &  $m~(\%)$          &   0.4  $\pm$     0.2    &   0.4  $\pm$      0.2   &   3.8  $\pm$  0.3    &  1.6  $\pm$    0.5  &     2.8  $\pm$    0.8  &    5.4  $\pm$   0.6   &    10.0  $\pm$   0.7   \\  
4.754  &  $P~\rm (mJy)$     &   0.90 $\pm$     0.54   &   0.83 $\pm$      0.52  &   5.41 $\pm$  0.44   &  1.75 $\pm$    0.42 &     1.08 $\pm$    0.41 &    3.76 $\pm$   0.41  &     6.23 $\pm$   0.41   \\  %
       &  $\chi~(^\circ)$   &  50.9  $\pm$    17.1    &  92.6  $\pm$     18.0   &  75.7  $\pm$  2.3    & 44.1  $\pm$    6.8  &    11.2  $\pm$   10.8  &   75.5  $\pm$   3.1   &    26.8  $\pm$   1.8   \\  %
       &  $m~(\%)$          &   0.4  $\pm$     0.2    &   0.4  $\pm$      0.2   &   4.4  $\pm$  0.3    &  1.8  $\pm$    0.5  &     1.9  $\pm$    0.7  &    5.6  $\pm$   0.6   &    11.5  $\pm$   0.7   \\  
4.974  &  $P~\rm (mJy)$     &   0.60 $\pm$     0.52   &   2.21 $\pm$      0.49  &   5.70 $\pm$  0.40   &  1.39 $\pm$    0.38 &     1.32 $\pm$    0.37 &    2.94 $\pm$   0.38  &     5.18 $\pm$   0.37   \\  %
       &  $\chi~(^\circ)$   &  38.0  $\pm$    24.6    &  65.7  $\pm$      6.3   &  68.9  $\pm$  2.0    & 42.1  $\pm$    7.8  &    21.4  $\pm$    8.0  &   74.8  $\pm$   3.6   &    25.7  $\pm$   2.0   \\  %
       &  $m(\%)$           &   0.3  $\pm$     0.2    &   0.9  $\pm$      0.2   &   5.0  $\pm$  0.3    &  1.7  $\pm$    0.5  &     2.5  $\pm$    0.6  &    4.5  $\pm$   0.5   &    10.0  $\pm$   0.7   \\  
5.055  &  $P~\rm (mJy)$     &   1.53 $\pm$     0.52   &   2.20 $\pm$      0.49  &   5.46 $\pm$  0.40   &  1.05 $\pm$    0.38 &    1.18  $\pm$    0.38 &    2.88 $\pm$   0.38  &     4.94 $\pm$   0.38   \\  %
       &  $\chi~(^\circ)$   &   2.2  $\pm$     9.7    &  40.7  $\pm$      6.4   &  58.8  $\pm$  2.1    & 30.2  $\pm$   10.4  &   10.1   $\pm$    9.1  &   70.4  $\pm$   3.7   &    23.8  $\pm$   2.1   \\  %
       &  $m~(\%)$          &   0.6   $\pm$     0.2    &   0.9  $\pm$      0.2   &   4.9  $\pm$  0.3    &  1.9  $\pm$    0.5  &    2.2   $\pm$    0.6  &    4.4  $\pm$   0.5   &     9.5  $\pm$   0.7   \\  
\hline Freq (GHz) &  & Core & X6 & X5  & X4 & X3 & X2 & X1\\
              (1) & (2) & (3) & (4) & (5) & (6) & (7) & (8) & (9) \\
8.120  &  $P~\rm (mJy)$     &   1.46 $\pm$   0.56  &   3.03 $\pm$  0.40   &   7.17 $\pm$   0.37  &  1.42 $\pm$   0.36  &    0.83 $\pm$   0.36 &    0.68  $\pm$  0.36 &    1.53 $\pm$  0.36   \\  %
       &  $\chi~(^\circ)$   &  72.4  $\pm$  10.9   & -55.6  $\pm$  3.7    &  -1.0  $\pm$   1.4   & 71.4  $\pm$   7.2   &   58.4  $\pm$  12.2  &  -71.8   $\pm$ 15.1  &   54.7  $\pm$  6.6   \\  %
       &  $m~(\%)$          &   0.5  $\pm$   0.2   &  2.4   $\pm$  0.3    &   9.7  $\pm$   0.5   &  4.3  $\pm$   1.1   &    4.2  $\pm$   1.7  &    1.7   $\pm$  0.8  &    8.5 $\pm$   2.3   \\  
8.200  &  $P~\rm (mJy)$     &   1.55 $\pm$   0.54  &  3.56  $\pm$  0.38   &   6.80 $\pm$   0.35  &  1.68 $\pm$   0.34  &    0.90 $\pm$   0.33 &    0.65  $\pm$  0.34 &    1.09 $\pm$  0.33   \\  %
       &  $\chi~(^\circ)$   &  65.1  $\pm$  10.0   &-59.8   $\pm$  3.0    &  -3.1  $\pm$   1.4   & 69.8  $\pm$   5.7   &   73.7  $\pm$  10.6  &  -74.9   $\pm$ 14.8  &   54.5  $\pm$  8.7    \\  %
       &  $m~(\%)$          &   0.5  $\pm$   0.2   &  2.9   $\pm$  0.3    &   9.2  $\pm$   0.5   &  5.2  $\pm$   1.0   &    4.7  $\pm$   1.7  &    1.7   $\pm$  0.8  &    6.4  $\pm$  2.2    \\  
8.400  &  $P~\rm (mJy)$     &   2.05 $\pm$   0.53  &  2.46  $\pm$  0.36   &   6.40 $\pm$   0.33  &  1.24 $\pm$   0.31  &    0.49 $\pm$   0.31 &     0.74 $\pm$  0.32 &    1.42 $\pm$  0.31   \\  %
       &  $\chi~(^\circ)$   &  60.0  $\pm$   7.4   &-66.1   $\pm$  4.1    &  -5.6  $\pm$   1.4   & 69.2  $\pm$   7.2   &   73.4  $\pm$  18.1  &   -75.2  $\pm$ 12.2  &   54.2  $\pm$  6.2    \\  %
       &  $m(\%)$           &   0.7  $\pm$   0.2   &  2.0   $\pm$  0.3    &   8.6  $\pm$   0.4   &  3.9  $\pm$   0.9   &    2.7  $\pm$   1.7  &     2.0  $\pm$  0.8  &    8.9  $\pm$  1.9    \\  
8.541  &  $P~\rm (mJy)$     &   1.73 $\pm$   0.54  &  2.43  $\pm$  0.37   &   6.78 $\pm$   0.34  &  1.58 $\pm$   0.33  &    0.69 $\pm$   0.33 &     0.44 $\pm$  0.33 &    1.17 $\pm$  0.33   \\  %
       &  $\chi~(^\circ)$   &  55.4  $\pm$   8.9   &-71.6   $\pm$  4.3    &  -7.6  $\pm$   1.4   & 67.5  $\pm$   5.9   &    52.0 $\pm$  13.5  &    -73.2 $\pm$ 21.4  &    53.3 $\pm$  8.0    \\  %
       &  $m~(\%)$          &   0.6  $\pm$   0.2   &  2.0   $\pm$  0.3    &   9.3  $\pm$   0.5   &  5.1  $\pm$   1.0   &     3.8 $\pm$   1.7  &      1.3 $\pm$  0.9  &     7.8 $\pm$  2.0    \\  
\hline Freq (GHz) &  & Core & U7  & U6 & U5 & U4 & U3\\
              (1) & (2) & (3) & (4) & (5) & (6) & (7) & (8) & (9) \\
15.144 &  $P~\rm (mJy)$     &  3.13  $\pm$  0.78    &  0.95  $\pm$     0.74 &   0.22 $\pm$  0.74  & 3.98  $\pm$ 0.74   &    1.91  $\pm$    0.74  &     1.10 $\pm$    0.74 &     ...        \\  %
       &  $\chi~(^\circ)$   &  74.5  $\pm$  7.1     & 70.8   $\pm$    22.3  &  93.8  $\pm$ 21.7   & 57.6  $\pm$ 5.3     &  141.5   $\pm$   11.0   &   122.7  $\pm$   19.2  &     ...        \\  %
       &  $m~(\%)$          &  1.8   $\pm$  0.4     &  2.5   $\pm$     1.9  &   0.6  $\pm$  2.1   & 17.3  $\pm$ 3.2    &   12.7   $\pm$    5.2   &    13.7  $\pm$    9.2  &     ...        \\  
15.224 &  $P~\rm (mJy)$     &  3.15  $\pm$  0.82    &  1.52  $\pm$     0.78 &   0.13 $\pm$  0.78  &  3.37 $\pm$ 0.78   &    1.55  $\pm$    0.78  &     1.24 $\pm$    0.78 &     ...        \\  %
       &  $\chi~(^\circ)$   & 92.8   $\pm$  7.4     & 75.7   $\pm$    14.6  &  63.8  $\pm$ 22.3   &  43.6 $\pm$ 6.6    &   145.0  $\pm$   14.3   &    135.8 $\pm$   17.9  &     ...        \\  %
       &  $m~(\%)$          &  1.3   $\pm$  0.4     &  4.0   $\pm$     2.0  &   0.4  $\pm$  2.1   & 14.0  $\pm$ 3.2    &    11.9  $\pm$    5.9   &    17.7  $\pm$   11.0  &     ...        \\  
15.425 &  $P~\rm (mJy)$     &  3.38  $\pm$  1.01    &  1.07  $\pm$     0.98 &   0.30 $\pm$  0.98  &  4.32 $\pm$ 0.98   &     1.03 $\pm$    0.98  &     1.55 $\pm$    0.98 &     ...        \\  %
       &  $\chi~(^\circ)$   & 95.1   $\pm$  8.5     &  63.1  $\pm$    14.6  &  75.7  $\pm$ 18.9   & 19.9  $\pm$ 6.5    &    12.9  $\pm$   21.7   &   132.0  $\pm$   18.0  &     ...        \\  %
       &  $m(\%)$           &  1.9   $\pm$  0.5     &  3.0   $\pm$     2.8  &   0.9  $\pm$  2.8   & 19.8  $\pm$ 4.2    &     8.0  $\pm$    7.5   &    17.2  $\pm$   10.8  &     ...            \\  
15.565 &  $P~\rm (mJy)$     &  3.58  $\pm$  1.15    &  1.17  $\pm$     1.12 &   0.54 $\pm$  1.12  &  4.17 $\pm$ 1.12   &     0.49 $\pm$    1.12  &    1.37  $\pm$    1.12 &     ...        \\  %
       &  $\chi~(^\circ)$   & 78.6  $\pm$  9.2     & 75.2  $\pm$    16.0  &  69.8 $\pm$ 19.4   & 20.4  $\pm$ 7.7    &    83.6  $\pm$   20.0   &   117.3  $\pm$   23.4  &     ...        \\  %
       &  $m~(\%)$          & 2.0    $\pm$  0.6     &   3.3  $\pm$     3.1  &    1.5 $\pm$  3.2   &  16.7 $\pm$ 4.4    &     4.1  $\pm$    9.3   &   17.2   $\pm$   14.0  &     ...        \\  
\hline
\end{tabular}
\begin{quote}
\ Notes: (1): Observing frequency. (2): Observed polarisation
information of every components, including polarised flux density,
polarised position angle and ratio of polarised flux to total flux
of components. (3)-(9): Component ID from core to jet, consistent
with the results from modelfitting.
\end{quote}
\end{scriptsize}
\end{table*}

\begin{table*} \caption{{\large Parameters in the depolarization analysis.}}
\begin{tabular}{lccccccccccccc}
\hline\hline
Component & $\rm ln~m_{0}$ (\%) & $RM$ ($\rm rad~m^{-2}$) & $b$ ($\rm \times 10^{5}~m^{-4}$) & $R$ \\ 
(1)       & (2)                 & (3)                         & (4)                              & (5) \\ 
\hline

C6  & -2.4 $\pm$ 0.4 & 1972 $\pm$ 207 & -1.9 $\pm$ 0.3  & -0.95 \\
C5  & -2.4 $\pm$ 0.1 &  722 $\pm$ 132 & -0.5 $\pm$ 0.1  & -0.96 \\
C4  & -3.7 $\pm$ 0.1 &  455 $\pm$  96 & -0.2 $\pm$ 0.1  & -0.70 \\
C3  & -4.0 $\pm$ 0.4 &  ...           &  0.2 $\pm$ 0.3  &  0.27 \\
C2  & -3.7 $\pm$ 0.1 &  152 $\pm$  43 &  0.5 $\pm$ 0.1  &  0.91 \\
C1  & -2.5 $\pm$ 0.2 &  113 $\pm$  36 &  0.2 $\pm$ 0.1  &  0.47 \\
X6  & -5.3 $\pm$ 0.5 & 2102 $\pm$ 124 &  8.8 $\pm$ 3.1  &  0.77 \\
X5  & -2.7 $\pm$ 0.2 &  780 $\pm$ 105 &  1.6 $\pm$ 1.3  &  0.50 \\
X4  & -2.9 $\pm$ 0.6 &  470 $\pm$ 127 & -0.8 $\pm$ 0.3  & -0.59 \\
X3  & -4.7 $\pm$ 0.9 &  ...           &  8.5 $\pm$ 5.2  &  0.50 \\
X2  & -5.0 $\pm$ 0.8 &    3 $\pm$ 235 &  5.0 $\pm$ 4.6  &  0.43 \\
X1  & -2.2 $\pm$ 0.6 &  176 $\pm$ 138 & -1.9 $\pm$ 1.8  & -0.20 \\

\hline
\end{tabular}
\begin{quote}
(1): Components name. (2): $\rm ln~m_{0}$ in \%. (3): Rotation
Measure in $\rm rad~m^{-2}$. (4): the slope $b$ in linear fitting.
(5): Pearson correlation coefficient for linear fitting.
\end{quote}
\end{table*}

\subsection{Physical Model-fitting}
In the framework of the relativistic beaming and the standard
synchrotron self-Compton (SSC) theory, the physical quantities in
the jets can be estimated under the assumption of homogeneous
spherically distributed emitting plasma (Ghisellini et al. 1993;
Readhead 1994) or an inhomogeneous relativistic jet model (Blandford
\& K$\rm \ddot{o}$nigl 1979 and K$\rm \ddot{o}nigl$ 1981). In this
work, we investigate physical model-fitting under the homogeneous
spherical assumption. Single-dish observations (Steppe et al. 1995,
Edwards \& Tingay 2004) indicate that the turnover frequency of
OQ172 as a
whole is located around $1\sim2$ GHz. 
The VLBA resolution at lower frequencies of 1.6 and 2.2 GHz is
insufficient to distinguish the core and inner jet components as
visible at the higher frequencies of 4.8, 8.3 and 15.3 GHz. In order
to analyse the radio spectrum for structures with similar sizes, we
combine the core and inner jet components into a core region
component at 4.8 (core+C6), 8.3 (core+X6) and 15.3 GHz (core+U7+U6)
with comparable size with the core at 1.6 and 2.2 GHz. The combined
component at 4.8, 8.3 and 15.3 GHz as well as the core component at
1.6 and 2.2 GHz are called hereafter the `core region component',
representing the total radio emission within 3 mas. The radio
spectrum of this core region component is constructed using its
total radio emission at each band, which clearly shows a turnover
(see Figure 7). In contrast, there is no turnover in the spectra of
other jet components. This strongly indicates that the SSA/FFA
absorption occurs in the core region (within 3 mas), not in the
outer regions.


Following O'Dea (1998), we apply both SSA and FFA absorption models
to fitting the spectrum of the core region component using formulas
presented by Kemeno et al. (2000):
\begin{equation}
S_{\nu}=S_{0}\nu^{2.5}[1-exp(-\tau_{s}^{\alpha-2.5})],
\end{equation}
and
\begin{equation}
S_{\nu}=S_{0}\nu^{\alpha}exp(-\tau_{f}\nu^{-2.1}),
\end{equation} where $\tau_{s}$ and $\tau_{f}$ are the SSA and FFA coefficients,
$S_{0}$, $\nu$ and $\alpha$ are the flux density at 1 GHz, the
frequency in GHz, and the spectra index, respectively.

\begin{table*} \caption{{\large Spectral fit for the
SSA and FFA models.}}
\begin{tabular}{lccccccccccccc}
\hline\hline Model & $S_{0}$ (Jy) & $\tau_{}$ & $\alpha$ & $\chi^{2}$\\ 
              (1) & (2) & (3) & (4) & (5) \\ 

\hline

SSA   & 0.25 $\pm$0.04 & 13.8 $\pm$2.9 ($\tau_{s}$) & 0.65  & 2.32 \\
FFA   & 1.63 $\pm$0.5  & 1.1 $\pm$0.4 ($\tau_{f}$)& 0.65 & 2.53  \\

\hline
\end{tabular}
\begin{quote}
(1): Absorbtion model. (2): $S_{0}$ in Jy. (3): Absorbtion
coefficient. (4): the spectral index. (5) $\chi^{2}$ for absorbtion
fitting.
\end{quote}
\end{table*}

Both SSA and FFA can well fit the spectrum of the core region
component (see Figure 7 and Table 4). However, the two models could
not be easily distinguished since the peak frequency is close to the
lowest frequency of our observation and due to the lack of
low-frequency observing data (i.e. at frequencies lower than the
turnover frequency). Future observations at lower frequencies would
help to resolve this ambiguity. While a specific model of absorption
cannot be determined yet, some conclusions on physical parameters in
the core region can be made. Based on the SSA model, we can obtain
the equipartition magnetic field $B$ $\sim3.4\pm0.6$ mG from the
relationship in O'Dea (1998)
\begin{equation}
\nu_{m}~\sim~8B^{1/5}S_{m}^{2/5}\theta^{-4/5}(1+z)^{1/5}\rm ~GHz,
\end{equation}
where the magnetic filed is in G, $S_{m}$ in Jy is a flux density at
a peak frequency $\nu_{m}$, $z$ is the redshift, $\theta$ is the
angular size in mas at $\nu_{m}$, which is taken as the core size at
1.6 GHz since the peak frequencies are very close to 1.6 GHz.

Alternatively, in the FFA model, the relationship between the FFA
coefficient of $\tau_{f}$ and the emission measure $\int^{L}_{0}
n_{e}^{2} dl$ can be expressed as (see, e.g., Gallimore, Elitzur \&
Baum 2006):
\begin{equation}
\tau_{f}=0.08235~\nu_{f}^{-2.1}T_{e}^{-1.35}\int^{L}_{0}
n_{e}^{2}dl,
\end{equation}
where $\nu_{f}$ is the peak frequency in GHz, $T_{e}$ is the
electron temperature in K, the electron density $n_{e}$ in $\rm
cm^{-3}$ and the path-length integral $l$ in pc, and $L$ is the FFA
path length. Assuming the equipartition brightness temperature of
$5\times10^{10}$ K (Readhead 1994) as intrinsic value of
$T_{B}^{'}$, our estimates of the brightness temperature (see Table
2) can be used to estimate the Doppler factor with
$\delta=T_{B}/T_{B}^{'}$ $\sim 10$ (Ghisellini et al. 1993). The
viewing angle then can be constrained as $\sin \theta\le 1/\delta$
(Ghisellini et al. 1993; Urry \& Padovani 1995). Assuming $\sin
\theta \sim 1/\delta$ and FFA happening in the core region component
with size of $\sim$ 3 mas (corresponding to a projected linear size
of $l\sim26.5$ pc), the path-length integral can be constrained as
$\int dl=l/\sin \theta\sim265$ pc along the line of sight. This size
is consistent with the NLR size estimated from $R_{NLR} = 295\times
L_{bol,46}^{0.47\pm0.13}$ pc (Mor et al. 2009) with the bolometric
luminosity $\log L_{bol}=46.10 \rm ~erg ~s^{-1}$. The average
electron density of the FFA absorber is
$n_{e}\sim10^{3}(T_{e}/10^{4})^{0.675} ~\rm cm^{-3}$, consistent
with typical values in NLR ($n_{e}\sim10^{3}~ \rm cm^{-3}$, and
$T_{e}\sim10^4$ K, Peterson 1997). This indicates that FFA is likely
caused by the NLR medium. Based on this estimated electron density,
and in combination with the relationship between RM, the electron
density and parallel magnetic filed (Equation 1), we can obtain the
magnetic filed $B_{\parallel}\sim 2.9$ mG by assuming RM in the core
region caused by the same plasma as FFA. In some GPS/CSS sources,
such as OQ 208, SSA and FFA might co-exist (e.g., Xie et al. 2005).
While our model-fitting does not allow us to distinguish between the
two models, we notice that the physical parameters calculated under
the assumption of a single absorption model (SSA or FFA) should be
treated as constrains, rather than real measurements.

\subsection{The Jet-NLR Interaction}

Based on the CIV line measurements from the SDSS spectrum, we
estimated the black hole mass $ M_{BH}=2.8\times10^{9}\rm M_{\odot}$
using the empirical relation by Kong et al. (2006)
\begin{equation}
M_{BH}(C_{IV})=4.6\times10^{5}(\frac{L_{CIV}}{10^{42}\rm
erg~s^{-1}})^{0.60\pm0.16} [\frac{\rm FWHM_{CIV}}{1000\rm
~km~s^{-1}}]^{2}\rm M_{\odot},
\end{equation}
The value is consistent within the error (typically, 0.5 dex) for a
black hole mass $8\times10^{9}\rm M_{\odot}$ estimated from the
$H\beta$ line, measured in the IR spectrum by Hirst et al. (2003).

Despite a low spectral resolution and blending with the nearby
$H\beta$ line, Hirst et al. (2003) found a very broad line width
$\sim2200~ \rm km~s^{-1}$ in $\rm [OIII]_{5007}$ line. Using the
$\rm [OIII]_{5007}$ line width as surrogate for the stellar velocity
dispersion $\sigma$ of the host galaxy $\sigma \sim$ $\rm FWHM
([OIII]_{5007}/2.35)$, Nelson 2000), the broad $\rm [OIII]_{5007}$
line makes OQ172 different from the $M_{BH} - \sigma$ relation of
local nearby galaxies (e.g., Tremaine et al. 2000)
\begin{equation}
 M_{BH}=10^{8.13} [\frac{\sigma}{200 \rm ~km~s^{-1}}]^{4.02}\rm M_{\odot},
\end{equation}
The large $\rm O[III]_{5007}$ line width could be due to the
interaction between the jet and NLR. This is supported by the recent
study of near-infrared spectra for a sample of powerful young radio
quasars by Kim et al. (2013), who found their $\rm [OIII]_{5007}$
lines are exceptionally broad, with FWHM $1300\sim2100 \rm ~km~
s^{-1}$, significantly larger than those of ordinary distant
quasars. They argued that these large line widths could be explained
by jet-induced outflows, as predicted by theoretical models of AGN
feedback.

All the available data that the jet bending and violent RM
variations occur at about 3 mas from the core. The Faraday
depolarisation appears to be caused by the internal medium in the
jet itself. The radio emission of the core region demonstrates a
spectrum with the turnover. Together with the broad $\rm
O[III]_{5007}$ line, all these properties are closely related and
likely caused by the same nuclei medium, presumably in the NLR.



\section{Conclusions}
We presented a VLBA polarimetry observation of the high-redshift GPS
quasar OQ172 at five frequency bands. Our conclusions are as
follows:
\begin{itemize}
\item{A core-jet morphology has been found at all five
frequencies with sharp bending at $\sim$ 3 mas from the core clearly
visible at 4.8, 8.3, and 15.3 GHz.}

\item{Linearly polarised emission has been detected in OQ172 at
all five frequencies, with a typical low fractional polarisation in
the core, and high in the jet components. The rotation measure
obtained at 4.8 and 8.3 GHz, shows the highest values of
$\sim2000~\rm rad~m^{-2}$ in the innermost region, dropping down to
$\sim700~\rm rad~m^{-2}$ at 3 mas, and decreasing to lower values
toward the outer jet regions.}

\item{The simultaneous multi-frequency observation enables us to
consider the turnover spectra within the framework of SSA and FFA
absorption models. While a specific model of absorption cannot be
determined yet, some conclusions on physical parameters in the core
region can be made.}

\item{A combination of the presented here and discussed observing
properties of the high-redshift GPS quasar OQ172 (the bending
structure of its VLBI jet, spectral turnover of compact components
and the broad $\rm O[III]_{5007}$) indicates on a possible role of
the same physical medium in the NLR.}




\end{itemize}

\section*{Acknowledgments}
We thank the anonymous referee for insightful comments and
constructive suggestions. We are grateful to  A.B. Fletcher for
helpful discussions and proofreading of the observing proposal. This
work is supported by the National Natural Science Foundation of
China (Grant No. 11233006, 11473069, 11273042, U1231106 and
U1431111). YL also acknowledges support from the Grant BK20131460.
MFG acknowledges support from the National Science Foundation of
China (grants 11473054 and U1531245) and by the Science and
Technology Commission of the Shanghai Municipality (14ZR1447100).
The VLBA is operated by the National Radio Astronomy Observatory
which is managed by Associated Universities, Inc., under cooperative
agreement with the National Science Foundation. The National Radio
Astronomy Observatory is a facility of the National Science
Foundation operated under cooperative agreement by Associated
Universities, Inc. This paper has made use of the NASA/IPAC
Extragalactic Database (NED), which is operated by the Jet
Propulsion Laboratory, California Institute of Technology, under
contract with the National Aeronautics and Space Administration.

{}

\newpage
\clearpage
\begin{figure}
  \begin{center}
    \mbox{
      \subfigure[]{{\includegraphics[height=0.35\textheight,angle=0]{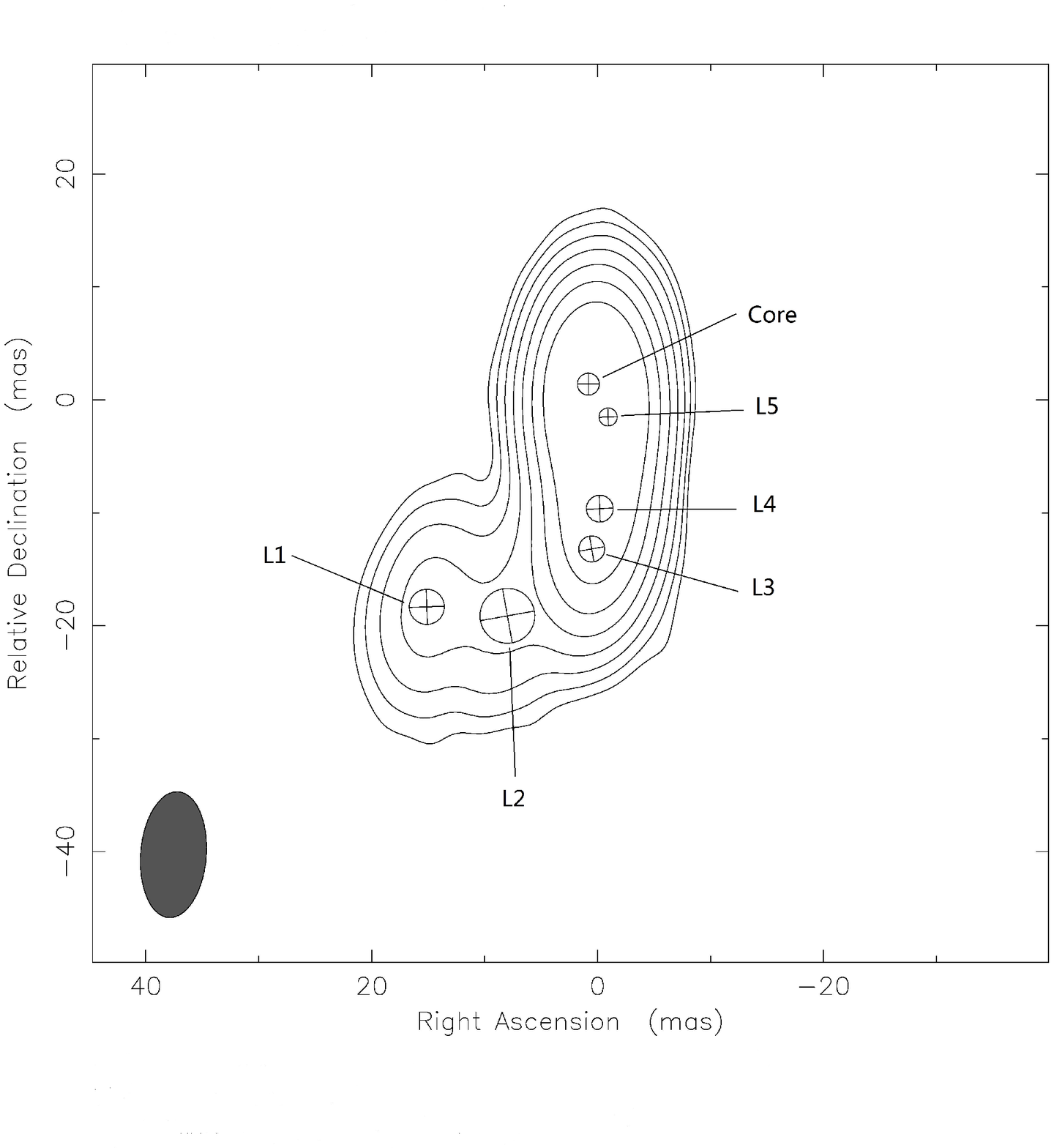}}}
      \quad
      \subfigure[]{{\includegraphics[height=0.35\textheight,angle=0]{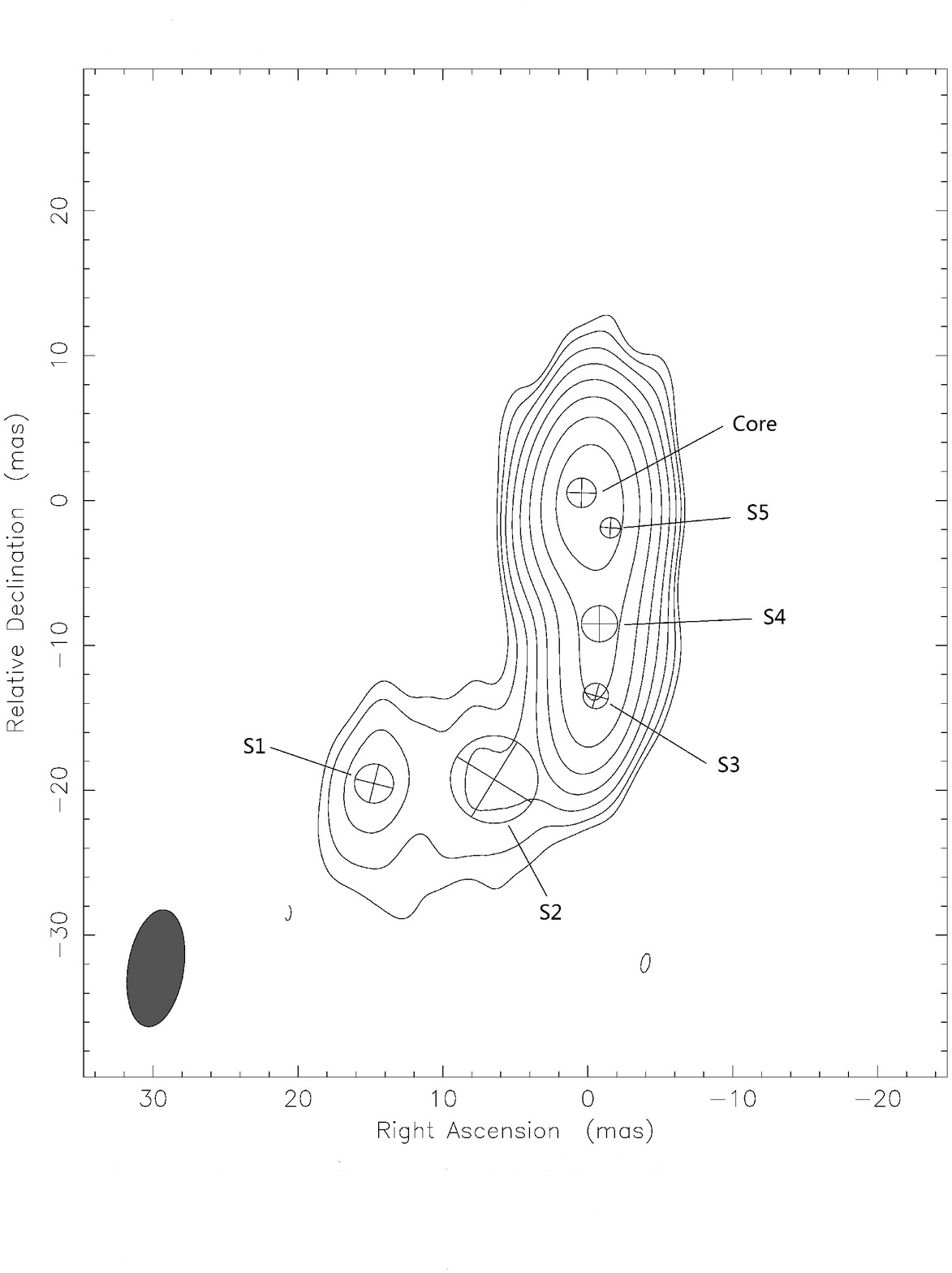}}}
      \quad
      \subfigure[]{{\includegraphics[height=0.35\textheight,angle=0]{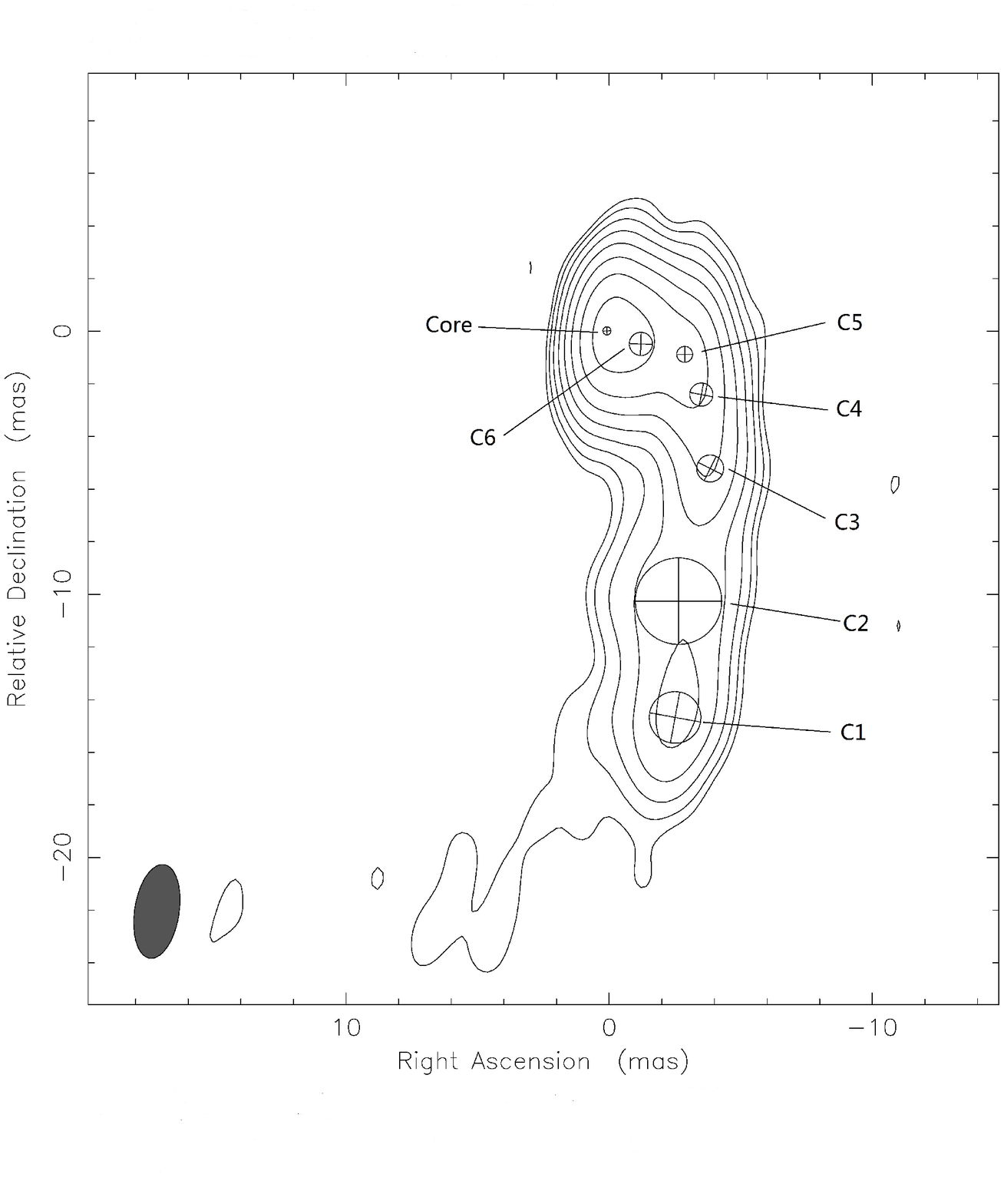}}}
      \quad}

\mbox{
      \subfigure[]{{\includegraphics[height=0.45\textheight,angle=0]{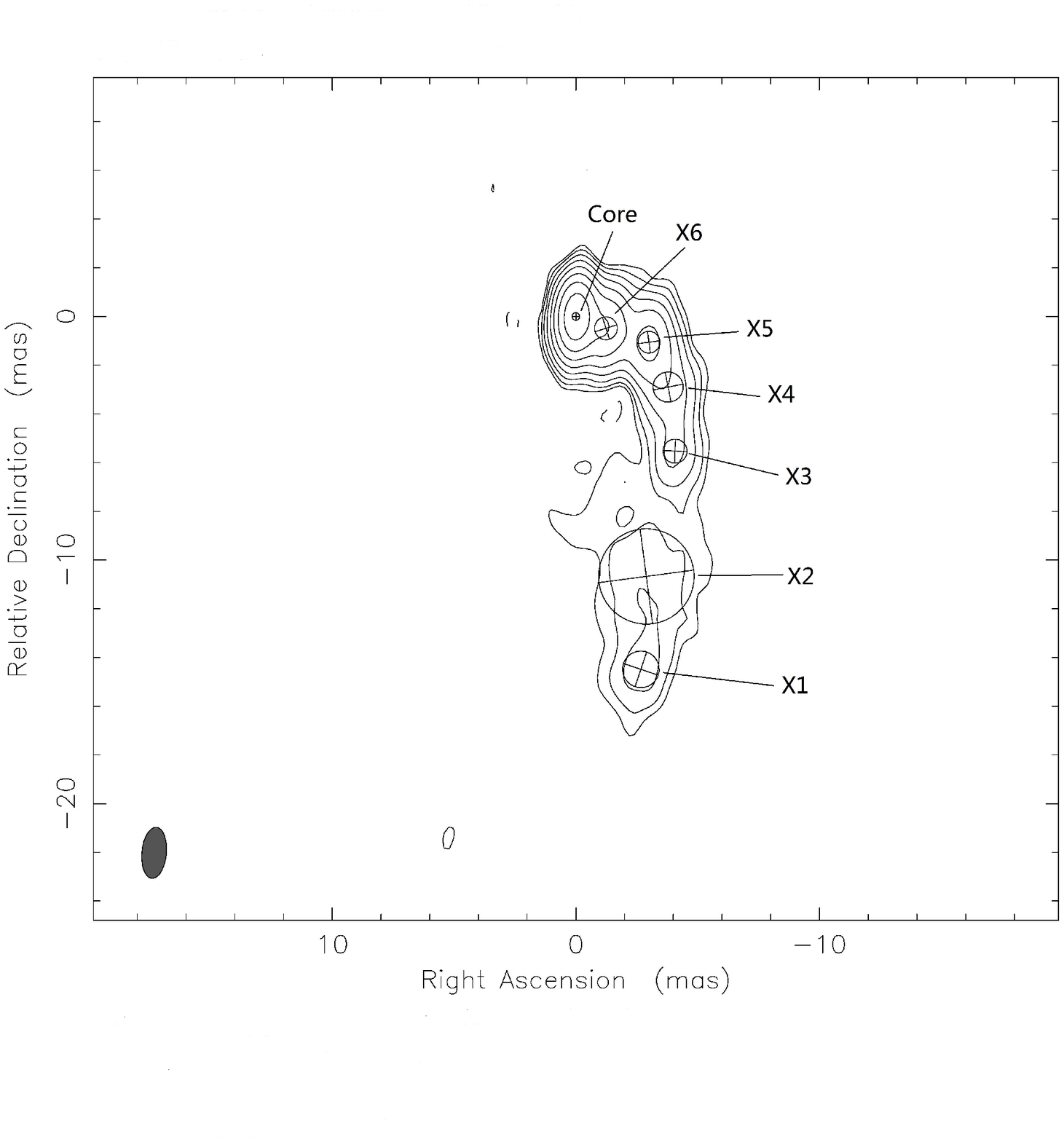}}}
      \quad
      \subfigure[]{{\includegraphics[height=0.45\textheight,angle=0]{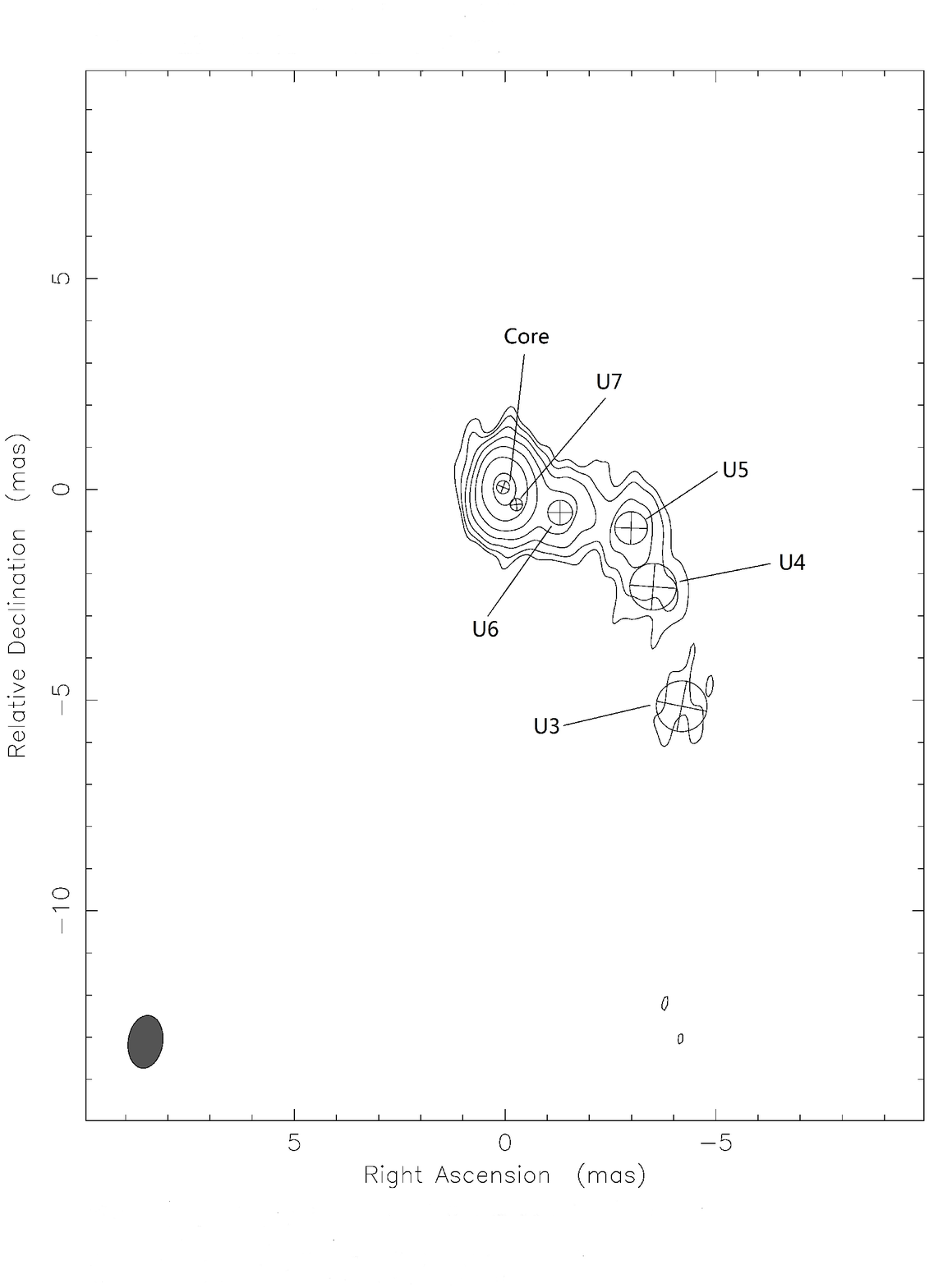}}}
      \quad    }
  \end{center}
  \caption{\bf Model fit to the total intensity visibilities at 1.6 GHz (a), 2.2 GHz (b), 4.8 GHz (c),
8.3 GHz (d) and 15.3 GHz (e) for the GPS quasar OQ172. The peaks and
beams are the same as Fig.2 - Fig.6 in Liu et al. (2016) for all
five frequencies. Each component is represented by an elliptic form
with the fitted position, size and orientation (see details in Table
2.).}
  \label{fig:critm0.05}
\end{figure}

\newpage
\clearpage
\begin{figure}
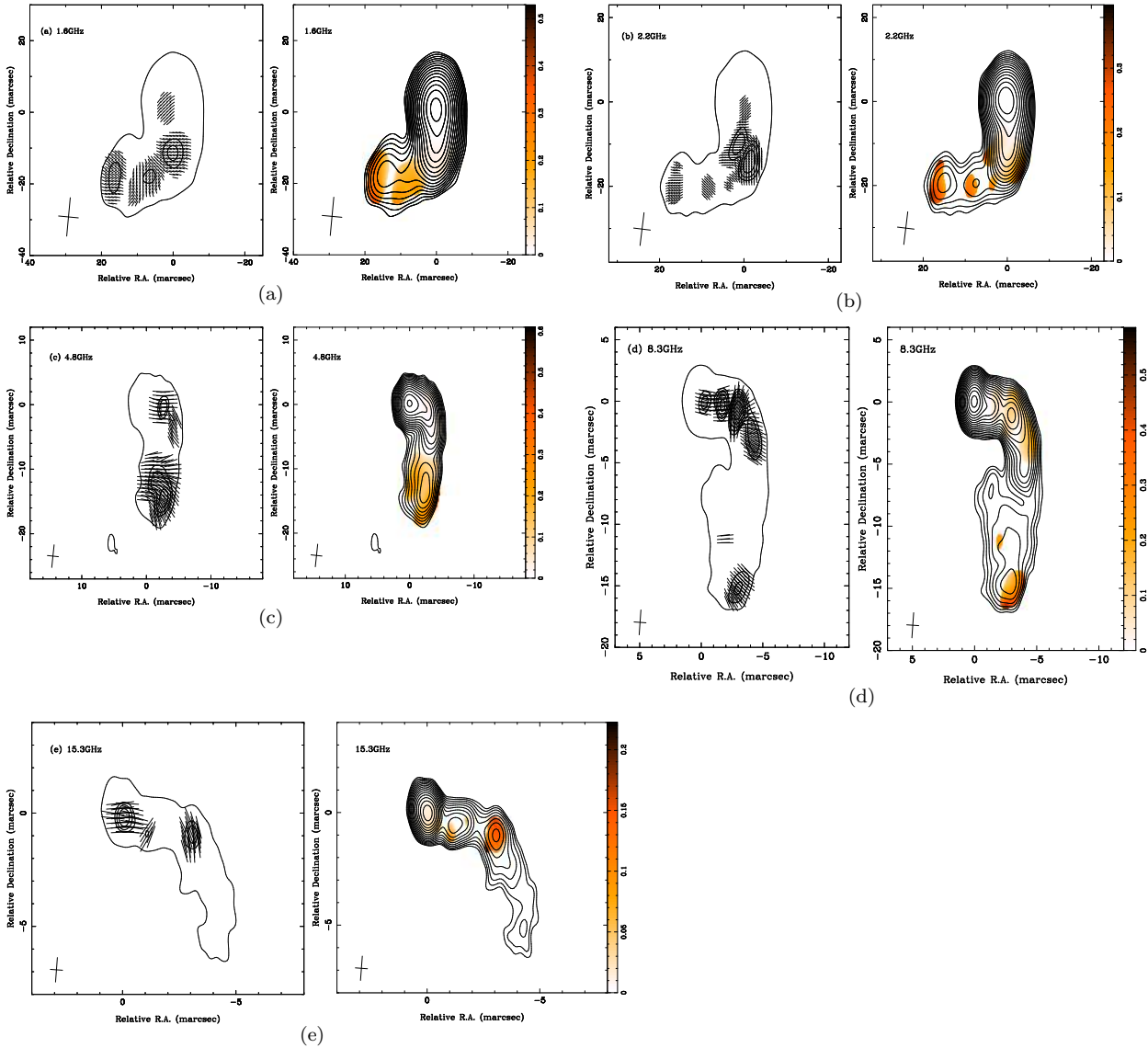

  \begin{center}
    \mbox{
      \subfigure[]{{
      \includegraphics[height=0.155\textheight,angle=270]{fig2a_1.eps}
      \includegraphics[height=0.165\textheight,angle=270]{fig2a_2.eps}
      }}
      \quad

      \subfigure[]{{
      \includegraphics[height=0.155\textheight,angle=270]{fig2b_1.eps}
      \includegraphics[height=0.165\textheight,angle=270]{fig2b_2.eps}
      }}
      \quad
    }

\mbox{
      \subfigure[]{{
      \includegraphics[height=0.155\textheight,angle=270]{fig2c_1.eps}
      \includegraphics[height=0.165\textheight,angle=270]{fig2c_2.eps}
      }}
      \quad

      \subfigure[]{{
      \includegraphics[height=0.16\textheight,angle=270]{fig2d_1.eps}
      \includegraphics[height=0.175\textheight,angle=270]{fig2d_2.eps}
      }}
      \quad
    }

    \mbox{
      \subfigure[]{{
      \includegraphics[height=0.18\textheight,angle=270]{fig2e_1.eps}
      \includegraphics[height=0.19\textheight,angle=270]{fig2e_2.eps}
      }}
      \quad
    }
 \end{center}
  \caption{\bf Polarisation distribution images for
the observation of GPS quasar OQ172 at five frequency bands. The
left image in each panel is an EVPA distribution (solid line) with
contours of continuum intensity overlaid; the right shows the
polarised intensity (color) overlaid by total intensity contours.
The lowest contours for total and linearly polarised intensity
distributions are 3 times the noise level. The color wedge at each
right image shows the fractional polarisation. (a): The restoring
beam has dimensions of 11.2 mas $\times$ 5.81 mas at position angle
$-4.47^{\circ}$, indicated at the bottom left corner. Contours start
at 4.21 $\rm mJy~beam^{-1}$ and increase by factors of 2. (b): The
restoring beam has dimensions of 8.15 mas $\times$ 3.15 mas at
position angle $-9.10^{\circ}$, indicated at the bottom left corner.
Contours start at 3.91 $\rm mJy~beam^{-1}$ and increase by factors
of 2. (c): The restoring beam has dimensions of 3.6 mas $\times$
1.69 mas at position angle $-8.79^{\circ}$, indicated at the bottom
left corner. Contours start at 1.64 $\rm mJy~beam^{-1}$ and increase
by factors of 2. (d): The restoring beam has dimensions of 2.1 mas
$\times$ 1.0 mas at position angle $-5.25^{\circ}$, indicated at the
bottom left corner. Contours start at 1.30 $\rm mJy~beam^{-1}$ and
increase by factors of 2. (e): The restoring beam has dimensions of
1.26 mas $\times$ 0.82 mas at position angle $-8.31^{\circ}$,
indicated at the bottom left corner. Contours start at 2.25 $\rm
mJy~beam^{-1}$ and increase by factors of 2.}
\end{figure}

\newpage
\clearpage
\begin{figure}
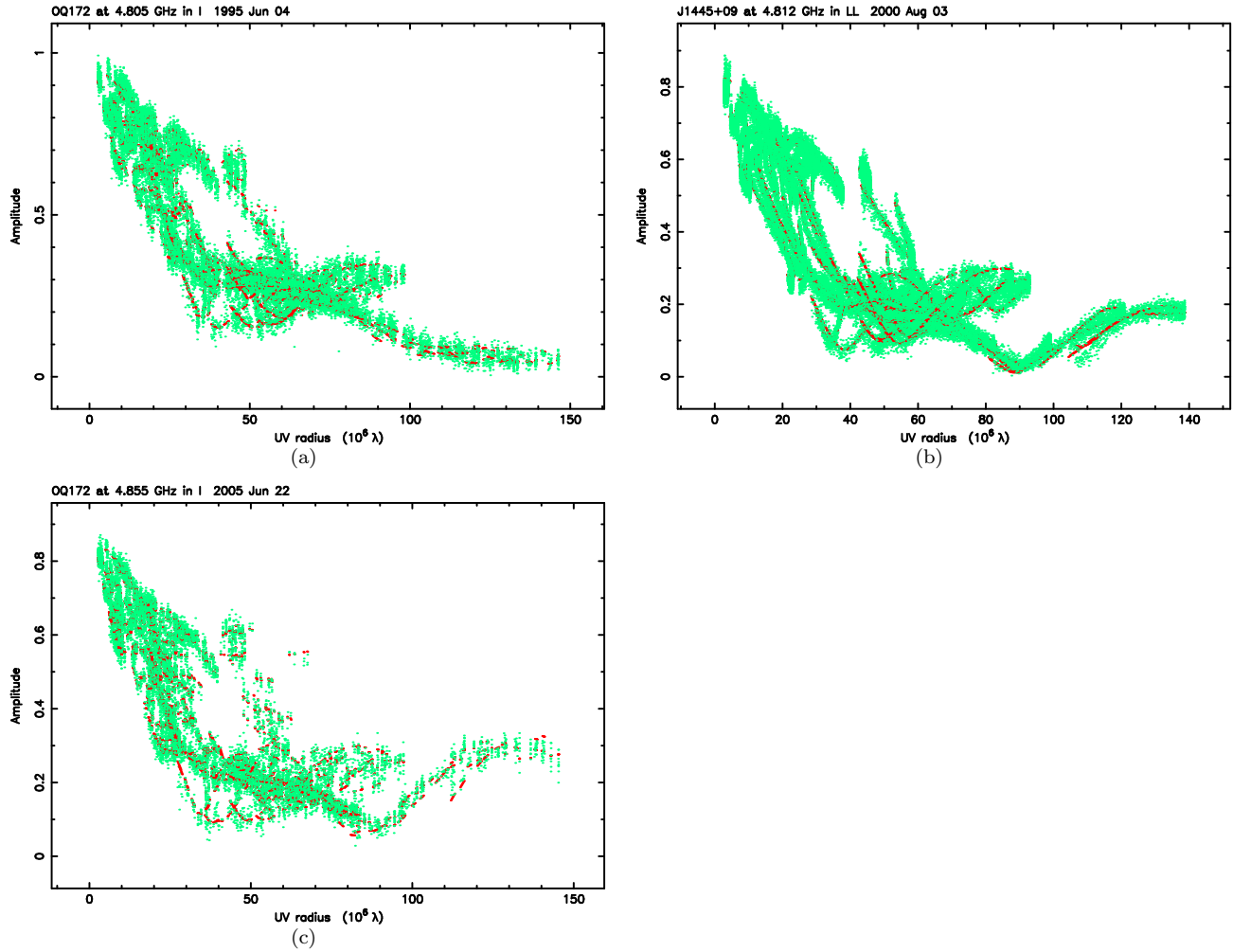

  \begin{center}
    \mbox{
      \subfigure[]{{\includegraphics[height=0.350\textheight,angle=270]{fig3a.eps}}}
      \quad
      \subfigure[]{{\includegraphics[height=0.350\textheight,angle=270]{fig3b.eps}}}
      \quad
    }
      \mbox{
      \subfigure[]{{\includegraphics[height=0.350\textheight,angle=270]{fig3c.eps}}}
      \quad
    }
  \end{center}
  \caption{\bf (a): The visibility amplitude for OQ172 observed at 5.0 GHz with VLBA on 1995.4 (Udomprasert's data, VLBA project code BT010);
  (b): The visibility amplitude for OQ172 observed at 5.0 GHz with VLBA on 2000.6 (VLBA project code W035);
  (c): The visibility amplitude for OQ172 observed at 4.8 GHz with VLBA on 2005.5 (our data in this work, VLBA project code BL129).}
  \label{fig:critm0.05}
\end{figure}

%


\newpage
\clearpage
\begin{figure}
  \begin{center}
    \includegraphics[height=0.9\textheight]{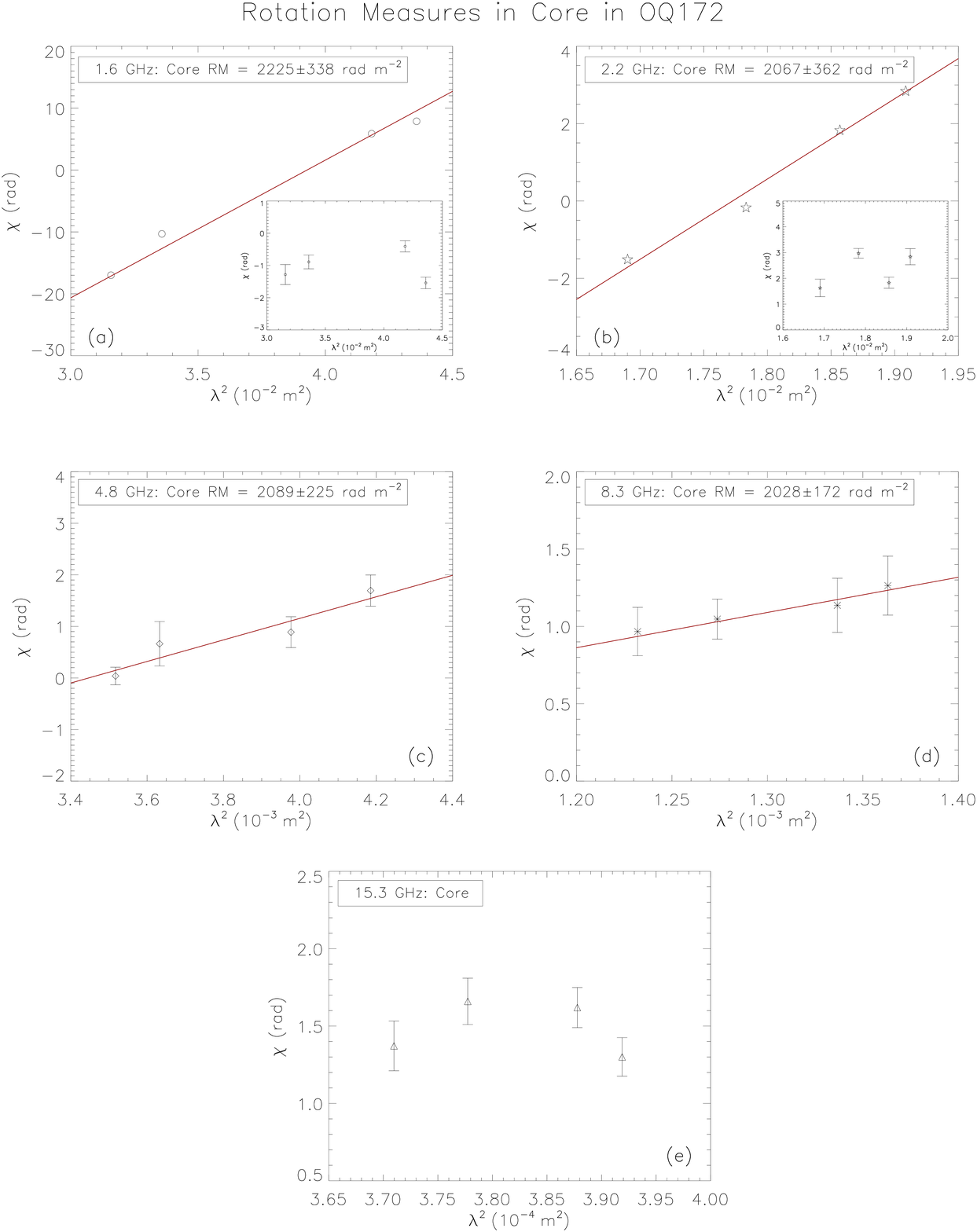}
  \end{center}
\caption{Rotation Measure for the quasar OQ172 in the core component
at five frequencies. The different kind of symbols are represented
different frequencies: cycles for 1.6 GHz, pentagrams for 2.2 GHz,
diamonds for 4.8 GHz, stars for 8.3 GHz, and triangles for 15.3 GHz,
respectively.}
\end{figure}

\newpage
\clearpage
\begin{figure}
  \begin{center}
    \includegraphics[height=0.9\textheight]{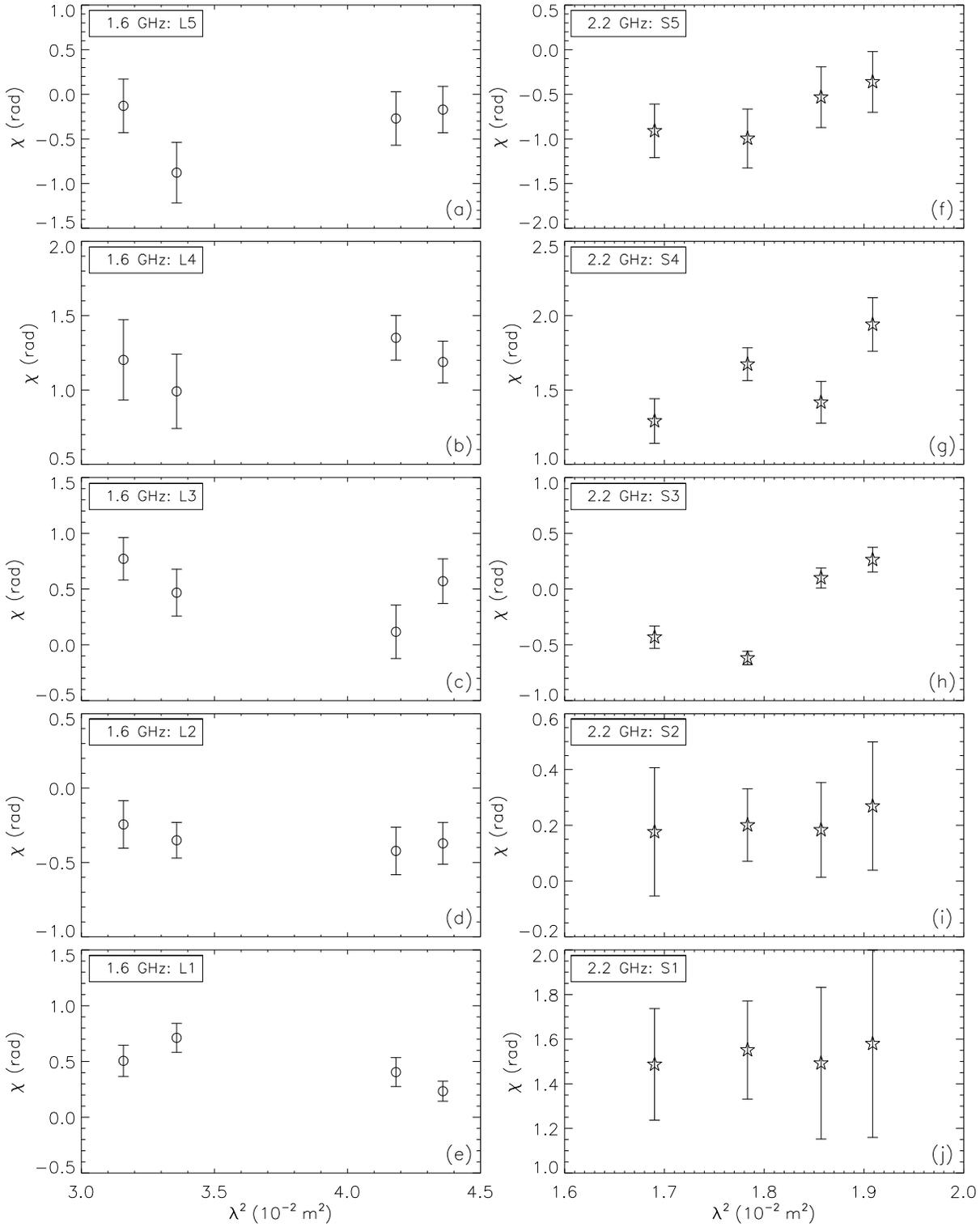}
  \end{center}
\caption{Rotation Measure for the quasar OQ172 in the jet components
at 1.6 and 2.2 GHz. The cycles show the components at 1.6 GHz and
the pentagrams display that at 2.2 GHz.}
\end{figure}


\newpage
\clearpage
\begin{figure}
  \begin{center}
    \includegraphics[height=0.9\textheight]{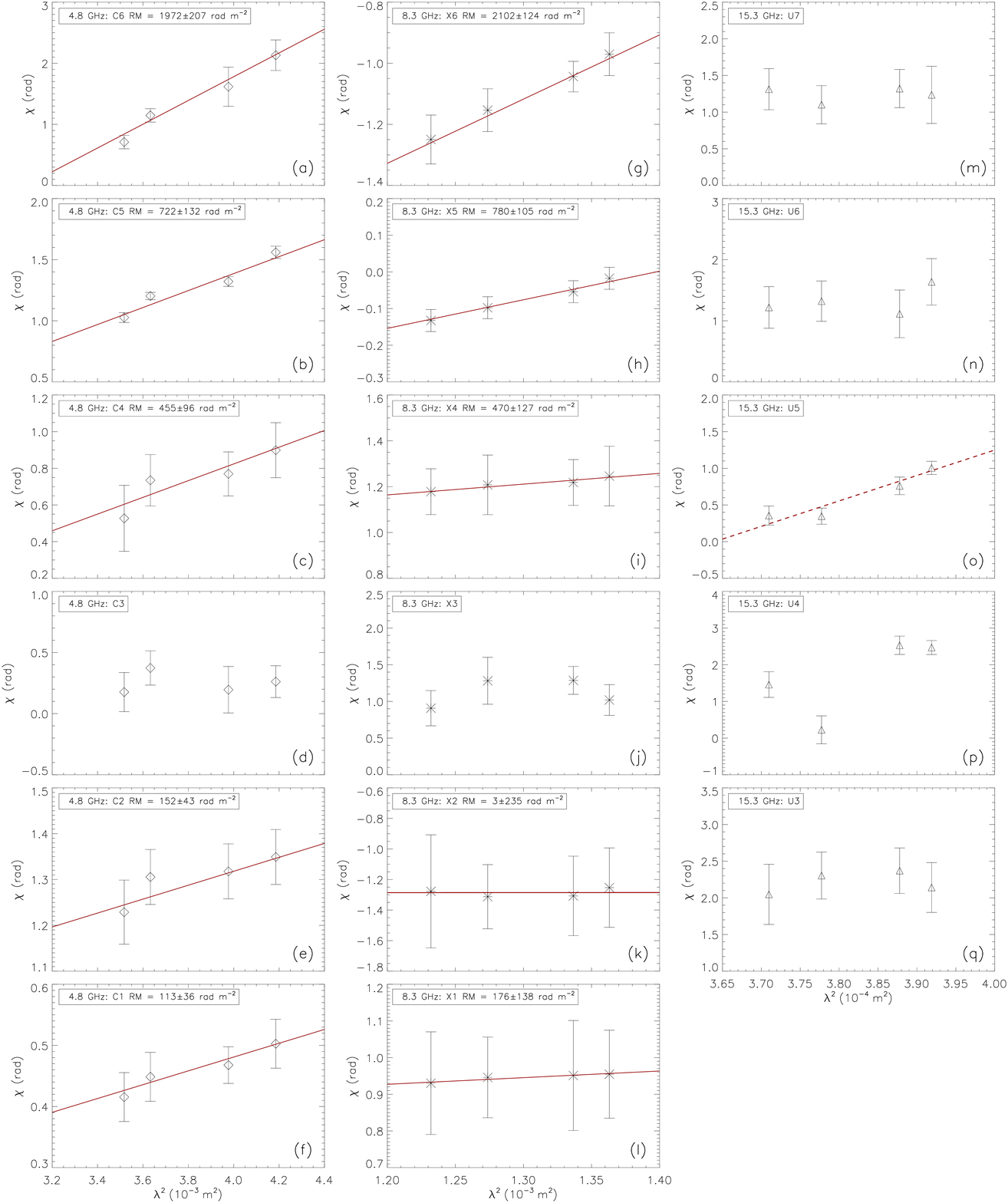}
  \end{center}
\caption{Rotation Measure for the quasar OQ172 in the jet components
at 4.8, 8.3 and 15.3 GHz. The diamonds present the components at 4.8
GHz, the stars describe the components at 8.3 GHz, and the triangles
indicate the components at 15.3 GHz, respectively.}
\end{figure}

\newpage
\clearpage
\begin{figure}
  \begin{center}
    \includegraphics[height=.5\textheight]{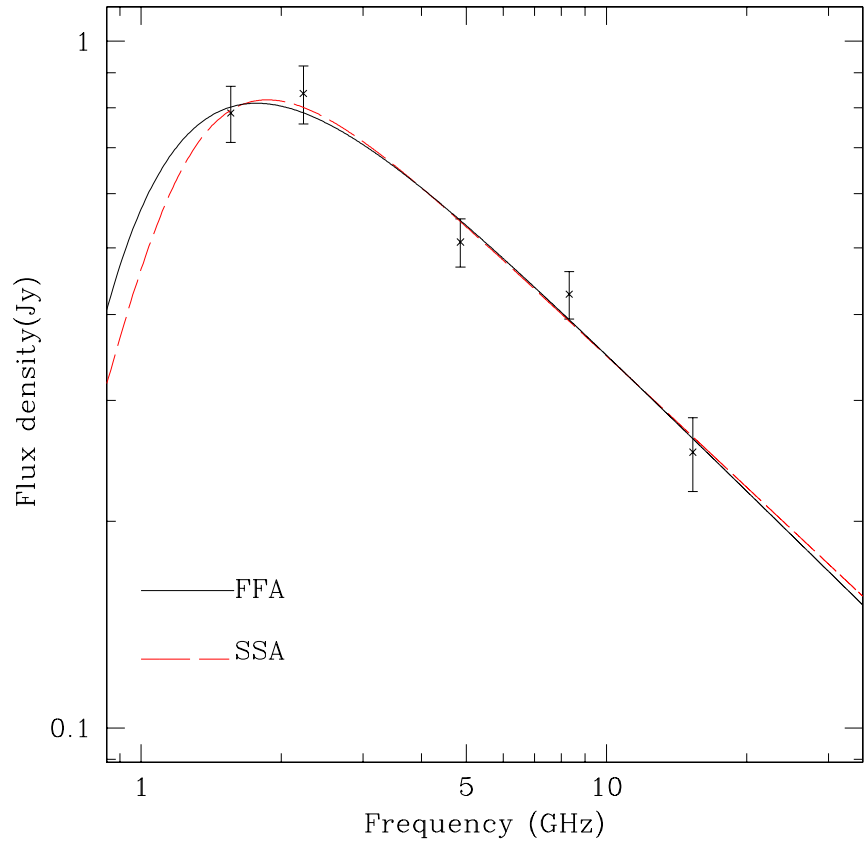}
  \end{center}
\caption{The core region component five-frequency spectrum from our
VLBA observation. The points with the error bars show the flux
density of the core region component, and the black line and red
dashed line indicate the fitted spectra for FFA and SSA models,
respectively.}
\end{figure}







\end{document}